\documentclass{emulateapj}

\usepackage{natbib}
\begin{document}

\title{IN-SYNC I: Homogeneous stellar parameters from high resolution APOGEE spectra for thousands of pre-main sequence stars}

\author{Michiel Cottaar\altaffilmark{1}, \and Kevin R. Covey\altaffilmark{2}, \and Michael R. Meyer\altaffilmark{1}, \and David L. Nidever\altaffilmark{3}, \and Keivan G. Stassun\altaffilmark{4}, \and Jonathan B. Foster\altaffilmark{5}, \and Jonathan C. Tan\altaffilmark{6}, \and S. Drew Chojnowski\altaffilmark{7}, \and Nicola da Rio\altaffilmark{6}, \and Kevin M. Flaherty\altaffilmark{8}, \and Peter M. Frinchaboy\altaffilmark{9}, \and Michael Skrutskie\altaffilmark{7}, \and Steven R. Majewski\altaffilmark{7}, \and John C. Wilson\altaffilmark{7}, \and Gail Zasowski\altaffilmark{7,10,11}}
\email{MCottaar@phys.ethz.ch}

\altaffiltext{1}{Institute for Astronomy, ETH Zurich, Wolfgang-Pauli-Strasse 27, 8093 Zurich, Switzerland}
\altaffiltext{2}{Lowell Observatory, Flagstaff, AZ 86001, USA}
\altaffiltext{3}{Department of Astronomy, University of Michigan, Ann Arbor, MI 48109, USA}
\altaffiltext{4}{Department of Physics \& Astronomy, Vanderbilt University, VU Station B 1807, Nashville, TN, USA}
\altaffiltext{5}{Yale Center for Astronomy and Astrophysics, Yale University New Haven, CT 06520, USA}
\altaffiltext{6}{Department of Astronomy, University of Florida, Gainesville, FL 32611, USA}
\altaffiltext{7}{Department of Astronomy, University of Virginia, Charlottesville, VA 22904, USA}
\altaffiltext{8}{Astronomy Department, Wesleyan University, Middletown, CT, 06459, USA}
\altaffiltext{9}{Department of Physics \& Astronomy, Texas Christian University, Fort Worth, TX 76129, USA}
\altaffiltext{10}{Department of Astronomy, The Ohio State University, Columbus, OH 43210, USA}
\altaffiltext{11}{Center for Cosmology and Astro-Particle Physics, The Ohio State University, Columbus, OH 43210, USA}

\begin{abstract}
Over two years 8,859 high-resolution H-band spectra of 3,493 young (1 - 10 Myr) stars were gathered by the multi-object spectrograph of the APOGEE project as part of the IN-SYNC ancillary program of that SDSS-III survey. Here we present the forward modeling approach used to derive effective temperatures, surface gravities, radial velocities, rotational velocities, and H-band veiling from these near-infrared spectra. We discuss in detail the statistical {\it and} systematic uncertainties in these stellar parameters. In addition we present accurate extinctions by measuring the E(J-H) of these young stars with respect to the single-star photometric locus in the Pleiades. Finally we identify an intrinsic stellar radius spread of about 25\% for late-type stars in IC 348 using three (nearly) independent measures of stellar radius, namely the extinction-corrected J-band magnitude, the surface gravity and the $R \sin i$ from the rotational velocities and literature rotation periods. We exclude that this spread is caused by uncertainties in the stellar parameters by showing that the three estimators of stellar radius are correlated, so that brighter stars tend to have lower surface gravities and larger $R \sin i$ than fainter stars at the same effective temperature. Tables providing the spectral and photometric parameters for the Pleiades and IC 348 have been provided online.
\end{abstract}


\maketitle
\section{Introduction}
\label{sec-1}
\label{sec:intro}
The dynamical state of young, recently formed stars places important constraints on the formation and early evolution of stars and star clusters. Hence this can help us to understand the complex interplay of turbulence, gravity, magnetic fields, and stellar feedback in star formation. Knowledge of the dynamical state of a stellar group as it emerges from its parental molecular cloud is also crucial for determining which star-forming regions will form bound clusters and which will disperse into the Galactic field population.

High-resolution spectra of molecular lines in the far-infrared and sub-mm have enabled the dynamics of the gas in molecular clouds to be studied with a typical precision of \(\sim 100\) m s\(^{-1}\). Unfortunately a similar precision is rarely reached for the young stars embedded in these molecular clouds. This is partly because these stars tend to be very faint in visible light due to the high extinction of the surrounding molecular cloud. This limitation can be overcome by studying these stars in the near-infrared, where radial velocity measurements are also less affected by the jitter due to stellar spots in the atmospheres of young stars \citep{Prato08, Huelamo08, Bailey12}. Although studies of high-resolution, near-infrared spectra have reached a precision of \(\sim 10\) m s\(^{-1}\) \citep[e.g.,][]{Bean10}, the lack of multi-object spectrographs in the near-infrared has limited these surveys to smaller sample sizes than collected by large visible-light spectroscopic surveys \citep[e.g.,][]{Tobin09, Gilmore12, Jeffries14}.

To fill this gap the Apache Point Observatory Galactic Evolution Experiment (APOGEE) project from the third Sloan Digital Sky Survey (SDSS-III; \citealt{Eisenstein11}) in 2012 approved the ancillary program "INfrared Spectra of Young Nebulous Clusters" (IN-SYNC) to take high resolution, near-infrared spectra of young stars in several star-forming regions with APOGEE multi-object spectrograph with the main goal of deriving the stellar dynamical state. So far, IN-SYNC has collected and analyzed more than 8,000 spectra of more than 3,000 young stars in IC 348, NGC 1333, NGC 2264, and the Orion A star-forming region. Here we discuss the spectral analysis for these young stars, including a discussion of the statistical and systematic errors. We provide the derived stellar parameters of IC 348 and the Pleiades \citep[the latter was observed by APOGEE prior to the approval of IN-SYNC and has already been released as part of the tenth data release;][]{Ahn14}. The dynamical state of the observed stars in IC 348 will be discussed in a companion paper (Cottaar et al., in prep).

In Section \ref{sec:obs} we briefly discuss the target selection, observations, and data reduction for the APOGEE IN-SYNC spectroscopic sample. In Section \ref{sec:spectra} we present our spectral analysis techniques, with estimates of the precision and accuracy of the resulting effective temperature, surface gravity, radial velocity, projected rotational velocity, and H-band veiling. We combine the spectroscopic effective temperature with 2MASS photometry to estimate the extinction and intrinsic luminosities of the observed stars in Section \ref{sec:phot}, which leads to a confirmation of a stellar radius spread in IC 348 in Section \ref{sec:lum_spread}. Finally we summarize our results in Section \ref{sec:conc}. Appendix \ref{app:tab} describes the accompanying online tables containing the stellar parameters derived in IC 348 and the Pleiades. The radial velocities are not included in this table, but will be released in a subsequent paper discussing the dynamics of IC 348 (Cottaar et al., in prep). The spectral parameters of the stars in NGC 1333, NGC 2264, and Orion A will be released in subsequent publications. All tables will be available through the IN-SYNC website (\url{http://www.astro.ufl.edu/insync/}).
\section{Observations}
\label{sec-2}
\label{sec:obs}
\subsection{Target selection}
\label{sec-2-1}
\label{sec:target}
The target selection for this spectroscopic survey in IC 348, NGC 1333, NGC 2264, and Orion A will be discussed in more detail in subsequent papers presenting analyses of each cluster's dynamics and stellar population. Here we provide a brief summary of the general approach. To maximize the efficiency of the program, potential targets were selected from existing catalogs of candidate or confirmed members. These catalogs include sources whose optical and/or near-infrared photometry is consistent with cluster membership, and exhibit other indicators of youth (infrared excess, X-ray activity, spectral signatures of accretion, low surface gravity, etc.).

\begin{figure}[htb]
\centering
\includegraphics[width=.9\linewidth]{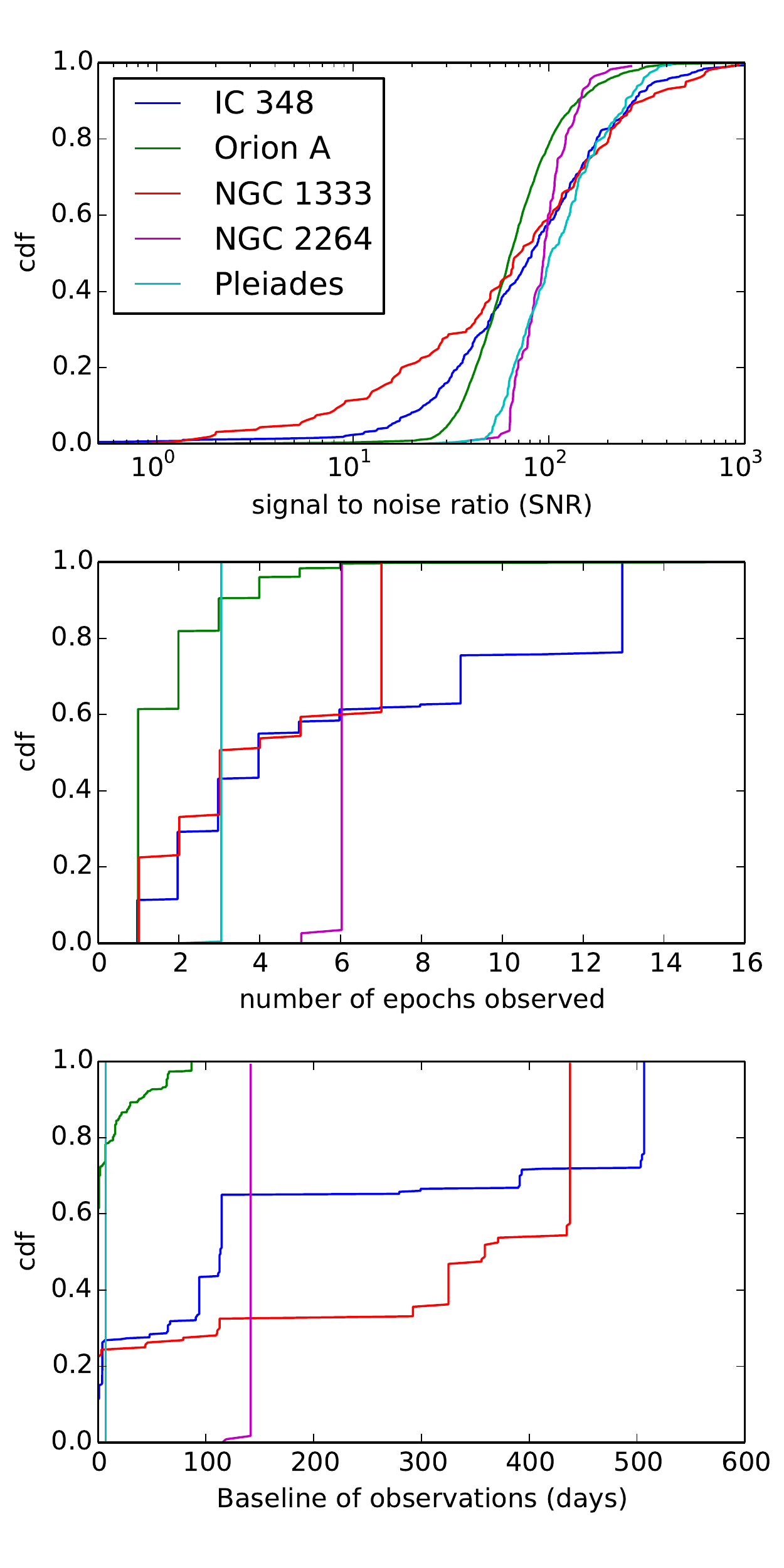}
\caption{\label{fig:observed}Cumulative distributions (cdf) of the S/N ratio of the observed spectra, the number of observed epochs, and the total baseline of those epochs for the 380 observed stars in IC 348 (blue), the 159 observed stars in NGC 1333 (red), the 115 observed stars in NGC 2264 (magenta), and the 2623 observed stars in Orion A (green).}
\end{figure}

IN-SYNC targets were selected from each cluster's membership catalog following a prioritization scheme designed to maximize the sample's utility for studying cluster dynamics. The IN-SYNC cluster targets have expected velocity dispersions of \(\sim 1\) km s\(^{-1}\), requiring precision of a few hundreds of m s\(^{-1}\) to resolve the cluster's internal velocity structure. We tested the S/N required to achieve sub-km/s RV precision using synthetic spectra of late-type (\(T_{\rm eff} = 3500-5000\) K), rapidly rotating (\(v \sin i \sim 20\) km s\(^{-1}\)), pre-main sequence (\(\log g = 4\)) stars, matched to APOGEE's wavelength coverage and resolution and masking spectral regions contaminated by bright OH night-sky emission lines. These tests indicated that RV precisions of \(\sim 100\) m s\(^{-1}\) could be achieved with S/N \(\sim\) 50 APOGEE spectra of late type stars, and in the absence of other astrophysical noise (e.g., spot effects). We thus assigned the highest priority to bright cluster members \((7.5 < \rm{H} < 12.5)\) for which S/N \(\sim\) 50 could be reached in a typical \(\sim 3\) hour integration time; fainter \((12.5 < \rm{H} < 15)\) cluster members were assigned lower priorities, and were only targeted if they did not conflict with brighter unobserved targets. The top panel of Figure \ref{fig:observed} shows the resulting distribution of the S/N.

The main limitation in observing all these targets was that APOGEE is unable to simultaneously observe stars within \(\sim 71.5\) arcseconds of each other due to collisions between the fibers. This means that even though multiple plates were drilled to cover most of the clusters, many stars in the dense cores could still not be observed. Multiple epochs were collected in all clusters to (1) identify close binaries and variability for stars outside of the crowded cluster centers and (2) increase the completeness of the observed stars in the dense stellar regions (Figure \ref{fig:observed}).

\subsection{APOGEE}
\label{sec-2-2}
\label{sec:apogee}
The spectra were collected with APOGEE's multi-object, high-resolution (\(R \sim 22,500\)) spectrograph with a spectral range covering much of the H-band from 1.51 to 1.69 \(\micron\) \citep{Wilson10, Wilson12}, which is fiber-fed from the Sloan 2.5m telescope \citep{Gunn06}. The instrument has been designed to carry out a spectrographic survey of \(\sim 100,000\) stars (mostly red giants) in the Milky Way (Majewski et al., in prep) as part of the SDSS-III survey. The light of up to 300 stars (of which \(\sim 230\) can be science targets) is collected by fibers connected to a pre-drilled plate. After dispersal with a transmissive Volume Phase Holographic grating, the 300 spectra are recorded by three H2RG near-infrared detectors. Both the reduced spectra and derived stellar parameters from APOGEE are regularly released as part of the SDSS-III data releases \citep[see][for a discussion of DR10]{Ahn14}. More information on those releases, the instrument, and the data reduction can be found on the SDSS-III web site (\url{http://www.sdss3.org}). 

In this work we analyze the reduced spectra made available in the single-epoch APOGEE apVisit FITS files. Nidever et al. (in prep) will completely describe the data reduction. In summary the correspondence between wavelength and pixel is computed separately for every fiber using exposures of ThArNe and UNe lamps with a final zero-point correction being applied to every spectrum based on observations of the bright telluric emission lines. The pipeline then subtracts the sky continuum and emission lines based on an interpolation of \(\sim 35\) sky fibers \citep{Zasowski13}. An additional \(\sim 35\) fibers are used to observe telluric (early-type) stars to get the spatial distribution of the telluric absorption spectra, allowing the absorption strength of each telluric species to be determined using model absorption spectra computed from the LBLRTM model atmosphere code\footnote{\url{http://rtweb.aer.com/lblrtm\_description.html}}. These species absorption strengths are then interpolated and the science spectrum is divided by the corresponding telluric spectrum. 

For most pixels the flux uncertainties have been computed as Poisson noise from the gain and readout noise in the observed flux, which are propagated throughout the pipeline. However, for pixels on telluric absorption or sky emission lines the statistical (but not systematic) uncertainties in the telluric corrections have also been propagated into the flux uncertainties (Nidever et al. in prep). There are two reasons why this noise estimate might underestimate the true flux uncertainties. Firstly, in the data reduction every spectrum is interpolated to a common wavelength grid, which leads to correlations in the flux uncertainties between neighboring pixels (i.e., red noise), which we do not take into account in the spectral fitting. Secondly, there might be additional sources of uncertainty related to, for example, sky subtraction, telluric correction, or a persistence in the observed flux sometimes seen between subsequent images \footnote{\url{http://www.sdss3.org/dr10/irspec/spectra.php\#errors}}. So we overestimate the information available in every spectrum to constrain the stellar parameters by ignoring both the correlation between the noise in neighboring pixels and these additional sources of uncertainty, which causes the stellar parameter uncertainties to be underestimated. In Section \ref{sec:stat} we will use the epoch-to-epoch variability to calibrate our underestimated stellar parameter uncertainties to the true uncertainties.
\section{Spectral analysis}
\label{sec-3}
\label{sec:spectra}
We apply a forward modeling approach to derive stellar parameters from the observed spectra, similar to that of \citet{Blake07, Blake10} and \citet{Bean10}. In Section \ref{sec:overview} we summarize this approach, with additional detail on the spectroscopic model given in Section \ref{sec:synthspec} and on the fitting procedure in Section \ref{sec:fit}. In Section \ref{sec:prec_acc} we discuss the precision and accuracy of the derived stellar parameters.

\begin{figure}[htb]
\centering
\includegraphics[width=.9\linewidth]{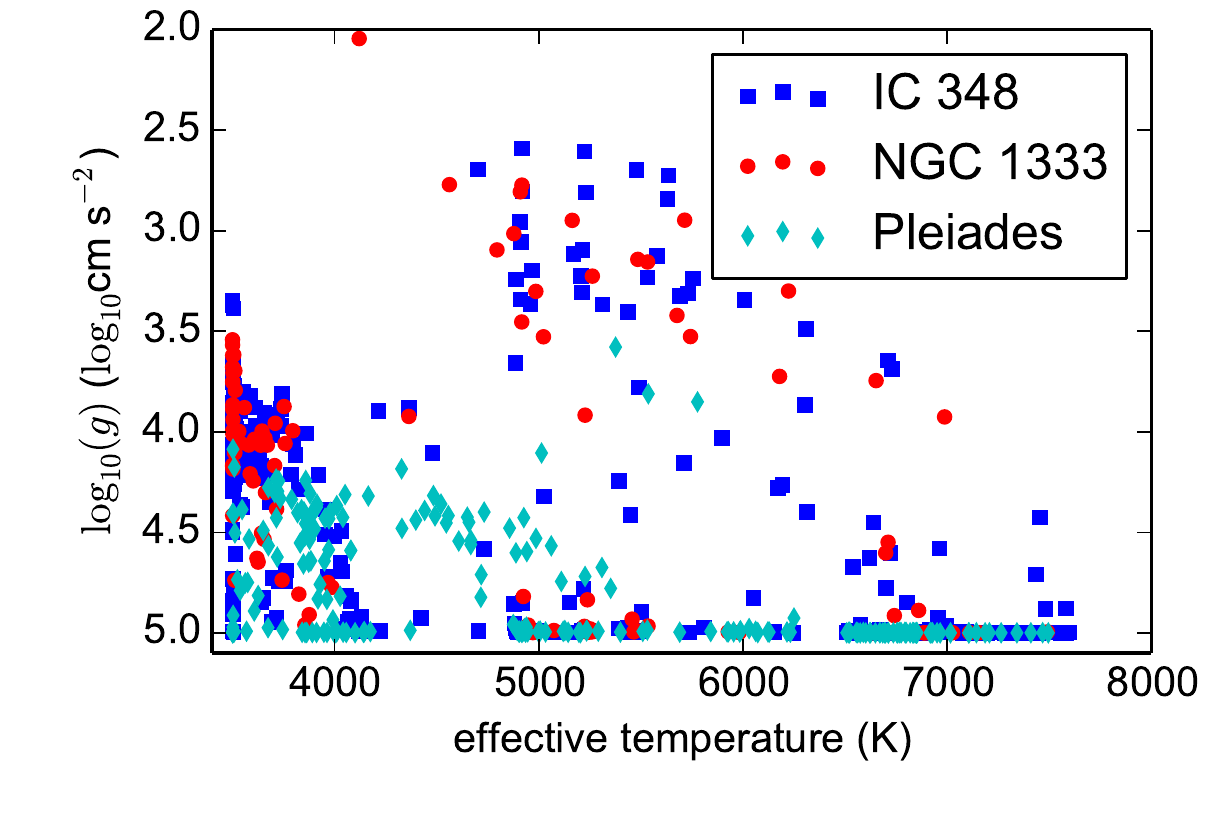}
\caption{\label{fig:aspcap}Effective temperature versus surface gravity as derived by ASPCAP (García Pérez et al. in prep) for stars in IC 348 (blue squares), NGC 1333 (red dots) and the Pleiades (cyan diamonds). The ASPCAP stellar parameters for these young stars pile up at the edge of the grid and do not form obvious sequence as expected. This in contrast to the stellar parameters derived from the analysis presented in this paper (see Figure \ref{fig:Teff_logg}).}
\end{figure}

This spectral analysis is separate from the APOGEE Stellar Parameters and Chemical Abundances Pipeline (ASPCAP; García Pérez et al. in prep). Using new ATLAS9 and MARCS model spectra \citep{Meszaros12} the ASPCAP pipeline is designed to derive accurate stellar parameters and chemical abundances for red giant stars \citep[e.g.,][]{Meszaros13, Smith13a}. Spectra of young stars tend to have different complications caused by emission lines, veiling, and stellar rotation, which have not been included in ASPCAP and cause it to produce inaccurate stellar parameters for these young stars (Figure \ref{fig:aspcap}). So the data analysis pipeline presented here was designed to fit the spectra of these young stars (while failing to reach the same accuracy as ASPCAP for red giants).

\subsection{Forward modeling}
\label{sec-3-1}
\label{sec:model}
\subsubsection{Overview}
\label{sec-3-1-1}
\label{sec:overview}
\begin{figure*}
\centering
\includegraphics[width=1\textwidth]{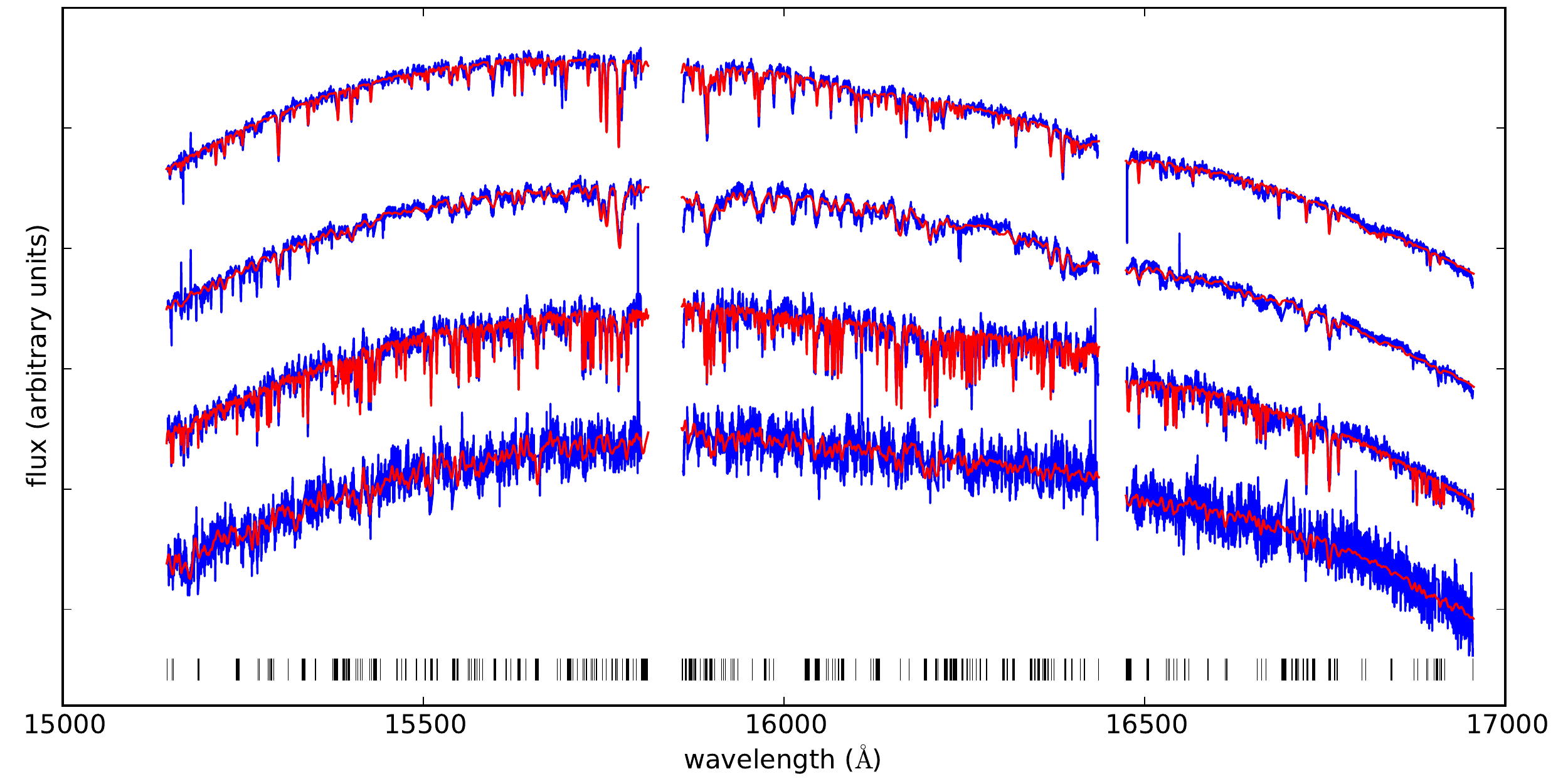}
\caption{\label{fig:sample}Sample fits of young stars in IC 348 with from top to bottom 2MASS J03442398+3211000 (\(\sim 6000\) K), 2MASS J03443916+3209182 (\(\sim 4500\) K), 2MASS J03445096+3216093 (\(\sim 3500\) K), and 2MASS J03425395+3219279 (\(\sim 2900\) K). The blue lines show one of the observed spectra for these stars and the red lines the best-fit model spectrum to each observed spectrum. Although high S/N spectra were selected as example, the S/N clearly increases towards the spectra of fainter and cooler stars at the bottom of the plot. Residual sky emission lines and bad pixels have been masked out, although some remnants can still be seen. The black lines at the bottom illustrate the typically masked regions. The two gaps in every spectrum are the gaps between APOGEE's three H2RG near-infrared detectors.}
\end{figure*}

We model the observed spectra with a grid of BT-Settl synthetic spectra \citep{Allard11} with five degrees of freedom: the effective temperature, surface gravity, radial velocity, rotational velocity, and H-band veiling due to possible continuum flux from a hot inner circumstellar disc. These synthetic spectra are then compared with the observed spectra and the \(\chi^2\) (over all good pixels as defined in section \ref{sec:fit}) is minimized. Finally the uncertainties on the best-fit stellar parameters are computed by a Markov chain Monte Carlo (MCMC) simulation. Sample fits are shown in Figure \ref{fig:sample}.

\subsubsection{The model}
\label{sec-3-1-2}
\label{sec:synthspec}
As a first step in our modeling process we prepared a grid of synthetic spectra. As a basis for our forward modeling approach, we adopt the grid of synthetic spectra from BT-Settl \citep{Allard11} at solar metallicity for the solar abundances from \citet{Caffau11} \footnote{This grid of synthetic spectra is available at \url{http://perso.ens-lyon.fr/france.allard/}.}. These spectra cover the complete visible and near-infrared wavelength range at very high resolution, and accurately reproduce interferometrically determined, bolometric effective temperatures when fit to low-resolution visible and near-infrared spectra of calibrator stars \citep{Mann13}. The grid spans from 400 to 7000 K with a step size of 100 K over the temperature range of interest (i.e. above 2100 K). The surface gravities range from -0.5 to 5.5 dex with a step size of 0.5 dex, although the exact range is temperature-dependent.

Prior to any fitting we degraded the BT-Settl spectra to match APOGEE's typical resolution of \(\sim 22,500\). To do this efficiently we resampled the spectra onto a wavelength grid covering the H-band with a sampling resolution of 700,000. The spectra were then convolved with a Gaussian, whose full width at half maximum is set to 700,000 / 22,500 = 31.1 pixels. Then we down-sampled the spectra to a sampling resolution of 300,000. This sampling resolution was chosen to minimize the computational time needed to rotationally broaden and interpolate the spectra, while still being sufficiently dense so that the spectrum can be accurately resampled to the observed wavelength grid using linear interpolation.

During the fitting procedure we vary the effective temperature and surface gravity as continuous variables. We use cubic interpolation to extract spectra at arbitrary effective temperatures and surface gravities between the grid points. These extracted spectra are then post-processed to produce synthetic spectra with different radial velocities, rotational velocities and levels of veiling:
\begin{enumerate}
\item The wavelength grid is Doppler-shifted by a given radial velocity.
\item The spectral lines are rotationally broadened by convolving the flux array with the rotational profile from equation 17.12 in \citet{Gray92}:
\begin{equation}
 {\rm kernel}(v_i) \propto 2 (1 - \epsilon) \sqrt{1 - \left(\frac{v_i}{v \sin i}\right)^2} + \frac{\pi \epsilon}{2} \left(1 - \left(\frac{v_i}{v \sin i}\right)^2\right),
\end{equation}
where $\epsilon=0.6$ is a parameter describing the limb darkening, $v \sin i$ is a variable representing the projected rotational velocity, and $v_i$ is the offset of pixel $i$ in units of velocity. For every pixel this kernel is integrated over the full width of the pixel, after which the sum of the kernel is normalized to one. For $v \sin i \approx 13$ km s$^{-1}$ the rotational broadening matches the broadening from the line-spread function (LSF) with a spectral resolution of 22,500. So any $v \sin i$ below this level will be difficult to detect and very sensitive to the line-spread function (see Section \ref{sec:vsini}).
\item A flat featureless flux is added to represent the veiling of the spectral lines due to flux from the inner part of the circumstellar discs. This veiling flux is parameterized by $R_{\rm H}$, the ratio of the non-stellar and stellar flux in the H-band. We advice to use this measure of the veiling with great caution, because in reality the flux from a circumstellar disc will have some wavelength dependency and this measure is also very sensitive to the exact subtraction of the sky continuum emission.
\end{enumerate}

Before comparing the fluxes from the synthetic and observed spectra, we fit the continua, so that the \(\chi^2\) is determined by how well the spectral lines match and is not dominated by the slope of the continuum. For this purpose we multiply the synthetic spectrum with a polynomial. This polynomial is recomputed for every synthetic spectrum to minimize the \(\chi^2\) of that fit. It is computed separately for the three chips of the APOGEE spectra. The degree of this polynomial is varied according to the optimization scheme described in section \ref{sec:fit}.
\subsubsection{The fitting}
\label{sec-3-1-3}
\label{sec:fit}
After matching the spectral continua, the \(\chi^2\) is computed for that synthetic spectrum. During this process any pixels covered by telluric emission lines or marked by the APOGEE pipeline as bad are masked: a typical spectrum has 20\% of its pixels masked (Figure \ref{fig:sample}).

The procedure described above allows us to compute the \(\chi^2\) between the synthetic and observed spectra for a set of stellar parameters efficiently. We minimize this \(\chi^2\) using a three-step procedure:
\begin{enumerate}
\item First an initial set of best-fit parameters are found for every spectrum by a global minimization of the \(\chi^2\) by varying the effective temperature, surface gravity, rotational velocity, radial velocities and veiling. At this stage a separate second-order polynomial is used to match the continuum of the observed spectrum in every chip. To find the global minimum we use the differential evolution routine implemented by Stepan Hlushak as part of the OpenOpt python library \citep{Kroshko07}. This routine alters a set of initially random candidate solutions by proposing for every candidate a new set of parameters by moving it in the direction of other candidate solutions. The proposed solution will replace the original solution only if it is an improvement in a \(\chi^2\)-sense. This optimization is run over the whole range of effective temperatures (100 to 7000 K) and surface gravities (0 to 5.5) covered by the BT-Settl grid, as well as a wide range in veiling (0 to 10\(^3\) times the continuum level), rotational velocity (0 to \(10^3\) km s\(^{-1}\)), and radial velocity (-10\(^3\) to 10\(^3\) km s\(^{-1}\)).
\item For the initial best-fit stellar parameters the degree of the polynomial used to match the continuum is optimized using the Bayesian Information Criterion (BIC; \citealt{Schwarz78}). According to BIC another parameter should only be added (by increasing the order of the polynomial) if it decreases the \(\chi^2\) with at least the natural logarithm of the number of data points (\(\ln n \approx 8\) for the APOGEE spectrum on a single chip). The optimum polynomial degree found in this step is typically between two and five and is kept fixed for each ship of every spectrum throughout the rest of the minimization and the MCMC simulation. The polynomial degree is capped at six to prevent the polynomial to correct for poorly fitted spectral features.
\item Finally, the \(\chi^2\) is minimized locally using a Nelder-Mead downhill routine to adjust the parameters in response to the change in the polynomial degree. During this final optimization the stellar parameters are only bounded by the limits in effective temperature and surface gravity of the grid of synthetic spectra, as well as the physical requirements of a non-negative veiling and rotational velocity. Typically the adjustments are small in this final step. The resulting minimum \(\chi^2\) is typically around a reduced \(\chi^2\) of \(1.5^2\) rather than 1, which we discuss in Section \ref{sec:stat}.
\end{enumerate}

Starting from the global \(\chi^2\) minimum we aim to compute the uncertainties on the best-fit stellar parameters. For this we use the MCMC routine "emcee" \citep{Foreman-Mackey13a}, which is based on the affine-invariant sampler from \citet{Goodman10}. In this routine multiple walkers (200 in our case) make a constrained random walk through the parameter space, where the direction of the next step is towards or away from the position of another random walker. The mean of this Markov chain is taken as the fiducial value. Although the distribution of the MCMC sample provides a first estimate of the statistical uncertainty, we will see in Section \ref{sec:stat} that this is a large underestimate of the true epoch-to-epoch variability. 
\subsection{Precision and accuracy}
\label{sec-3-2}
\label{sec:prec_acc}
\subsubsection{Statistical uncertainties on stellar parameters}
\label{sec-3-2-1}
\label{sec:stat}
\begin{figure}[htb]
\centering
\includegraphics[width=1\linewidth]{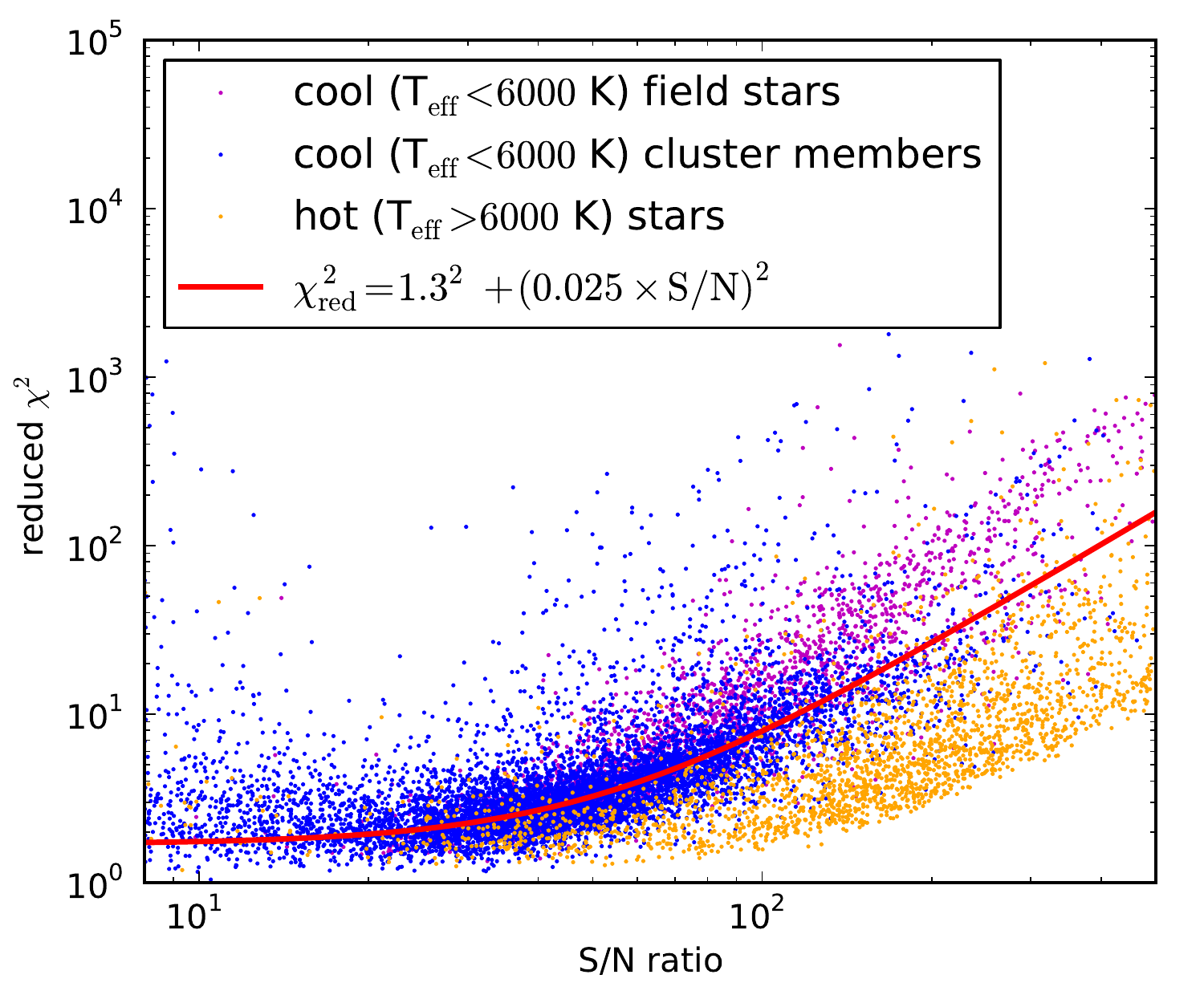}
\caption{\label{fig:redchisq}Reduced \(\chi^2\) of the best-fit model plotted against the S/N for all analyzed spectra. The red line shows simple model for the reduced \(\chi^2\) based on equation \ref{eq:chired} with an underestimate in the error array of 1.3 and an average systematic offset between observed and model spectrum of 2.5\% per pixel (i.e., \(\Delta_{\rm mod} = 0.025\)), appropriate for the vast majority of cluster members (in blue).}
\end{figure}
For non-variable stars the uncertainties given by the MCMC should match the variability in the stellar parameters measured for the same star at different epochs. Here we will compute this epoch-to-epoch variability to show that the MCMC uncertainties are probably an underestimate of the true uncertainty. We estimate the ratio of the true and MCMC uncertainty \(\eta_i\) for every epoch \(i\) by computing:
\begin{equation}
  \eta_i = \frac{p_i - \overline{p}}{\sqrt{\sigma_i^2 + \sigma_{\overline{p}}^2}}, \label{eq:eta}
\end{equation}
where \(p_i\) and \(\sigma_i\) are the observed stellar parameter and its measurement uncertainty from the MCMC at epoch \(i\) and \(\overline{p}\) and \(\sigma_{\overline{p}}\) are the weighted mean and its uncertainty computed over all other epochs (i.e., excluding epoch \(i\)). For a target with only two epochs this corresponds to simply computing the difference between the stellar parameters measured in the two epochs normalized by the quadratic sum of the uncertainties. For targets with more epochs, \(\eta_i\) calculates for every epoch the offset from the weighted mean of the other epochs, normalized again by the quadratic sum of the uncertainties. In this section we study the trend of \(\eta_i\) to show that the MCMC uncertainties underestimate the true variability due to in part the real flux uncertainty being about 30\% higher than estimated by the data reduction pipeline in most spectra with a larger correction needed for a small subset of spectra.

To estimate by which factor the noise is underestimated we employ the reduced chi-squared. The reduced \(\chi^2\) can be understood as the sum of two components: the contribution from the random noise in the flux values, which should be very close to one assuming an accurate representation of the noise in the spectrum, and the contribution from a systematic offset between the models and the observed spectra. In equation form this gives
\begin{equation}
\chi^2_{\rm red} = \chi^2_{\rm red,\ cont} + {\rm (S/N)}^2 \Delta_{\rm mod}^2, \label{eq:chired}
\end{equation}
where \(\chi^2_{\rm red,\ cont}\) is the reduced chi-squared for a perfect model, S/N is the signal to noise ratio, and \(\Delta_{\rm mod}\) is the average systematic offset (relative to the flux level) between the best-fit model and the observed spectrum. Figure \ref{fig:redchisq} shows that for the vast majority of stars in IC 348, NGC 1333, NGC 2264, Orion A and the Pleiades the reduced \(\chi^2\) indeed follows the trend with S/N expected from equation \ref{eq:chired}, which tells us that there is a systematic offset between the observed and the model spectra of about 2.5\% of the flux level for these young stars. Only for the hottest stars in these clusters (\(T_{\rm eff} > 6000\) K) do we find a somewhat smaller systematic offset. This allows us to define a reduced \(\chi^2\) corrected for they systematic differences between model and observations:
\begin{equation}
\chi^2_{\rm red,\ cont} = \chi^2_{\rm red} - 0.025^2 {\rm (S/N)}^2, \label{eq:chicorr}
\end{equation}
which is appropriate for all cool cluster members. This corrected reduced \(\chi^2\) is a measure of how much the noise is underestimated in the spectrum and should be close to one if the noise array is accurate.

\begin{figure}[htb]
\centering
\includegraphics[width=\linewidth]{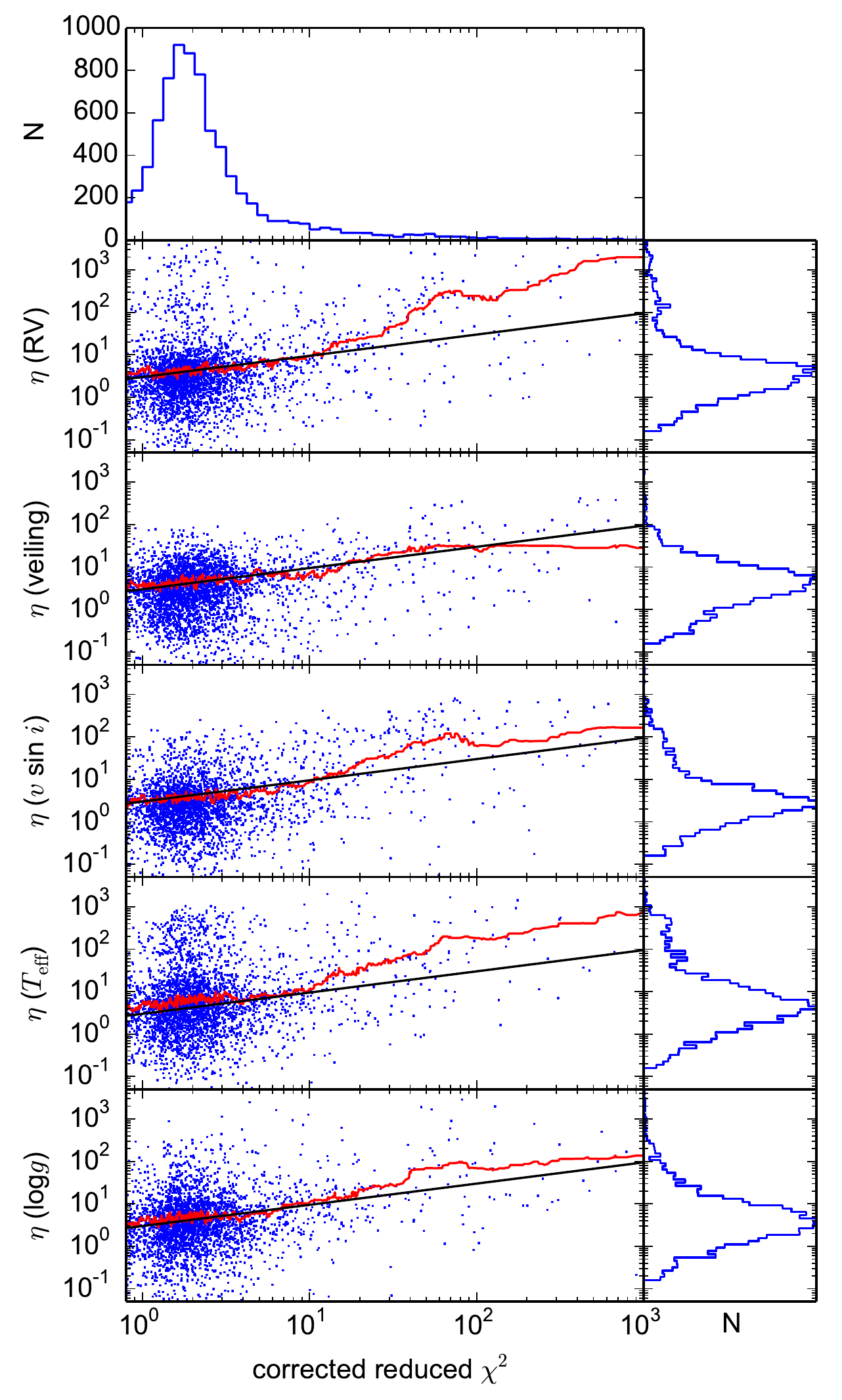}
\caption{\label{fig:stat}On the y-axis is plotted the epoch-to-epoch variability as measured by the offset in the measured stellar parameter in a specific epoch relative to the mean stellar parameter measured for that star over all other epochs. This offset is divided by the MCMC uncertainty (see equation \ref{eq:eta}). \(\eta\) is computed for the radial velocity (RV), the veiling, the rotational velocity (\(v \sin i\)), the effective temperature (\(T_{\rm eff}\)) and the surface gravity (\(\log g\)). On the x-axis the reduced \(\chi^2\) corrected for the systematic offset between best-fit model and observation (equation \ref{eq:chicorr}). Every dot represents a spectrum observed for a star in one of the analyzed clusters (IC 348, NGC 1333, NGC 22674, Orion A, and the Pleiades) excluding likely background stars. Histograms of these distributions have been plotted on the top and right. The red line shows the trend line derived by median-filtering over 100 stars with similar reduced \(\chi^2\). The black line represents the expected trend if the larger reduced \(\chi^2\) in the continuum is due to an underestimate of the flux uncertainty in these spectra \emph{and} if the true uncertainties are 2.5 times higher than expected from the flux uncertainty alone.}
\end{figure}

Now we can compare whether our underestimation of the epoch-to-epoch variability (as estimated from \(\eta_i\) from equation \ref{eq:eta}) is correlated with our underestimation of the flux uncertainty (from the corrected \(\chi^2_{\rm red}\) from equation \ref{eq:chicorr}). For all five stellar parameters \(\eta_i\) is measured at every epoch for every star with two or more spectra. The absolute value of \(\eta_i\) is plotted in the scatter plots of Figure \ref{fig:stat} against the reduced \(\chi^2\) corrected for the systematic differences between model and observations (equation \ref{eq:chicorr}). The red lines show the trend of \(\eta_i\) as obtained from median filtering and multiplying the median by 1.4826. The multiplicative factor of 1.4826 was chosen so that the red line now illustrates by how much the MCMC uncertainties underestimate the epoch-epoch variability at every reduced \(\chi^2\) corrected for the systematic offset between model and observed spectrum. If the MCMC uncertainties perfectly predict the epoch-to-epoch variability the red line would only show small fluctuations around one. The increase of this trend line towards the right in Figure \ref{fig:stat} suggests that the flux uncertainty in the spectra is underestimated by the square root of the corrected reduced \(\chi^2\). So we can calibrate the MCMC uncertainty to match the epoch-epoch variability to get the true measurement uncertainty \(\sigma_{\rm meas}\):
\begin{equation}
  \sigma_{\rm meas} = 3 \sigma_{\rm MCMC} \sqrt{\chi^2_{\rm red,\ cont}} \label{eq:sigtrue}
\end{equation}
where \(\sigma_{\rm MCMC}\) is the uncertainty given by the MCMC and \(\chi^2_{\rm red,\ cont}\) is given by equation \ref{eq:chicorr}. The factor of 3 represents an increase in the true uncertainties, even after correcting for underestimating the flux uncertainty and has been set to match the observed trend of the MAD in Figure \ref{fig:stat} by eye. This could be caused by correlation in the noise level between pixels, non-Gaussian errors in the flux and/or spectral line variability. We use these increased uncertainties throughout this paper and they are also provided in the online tables.

\begin{figure*}
\centering
\includegraphics[width=1\textwidth]{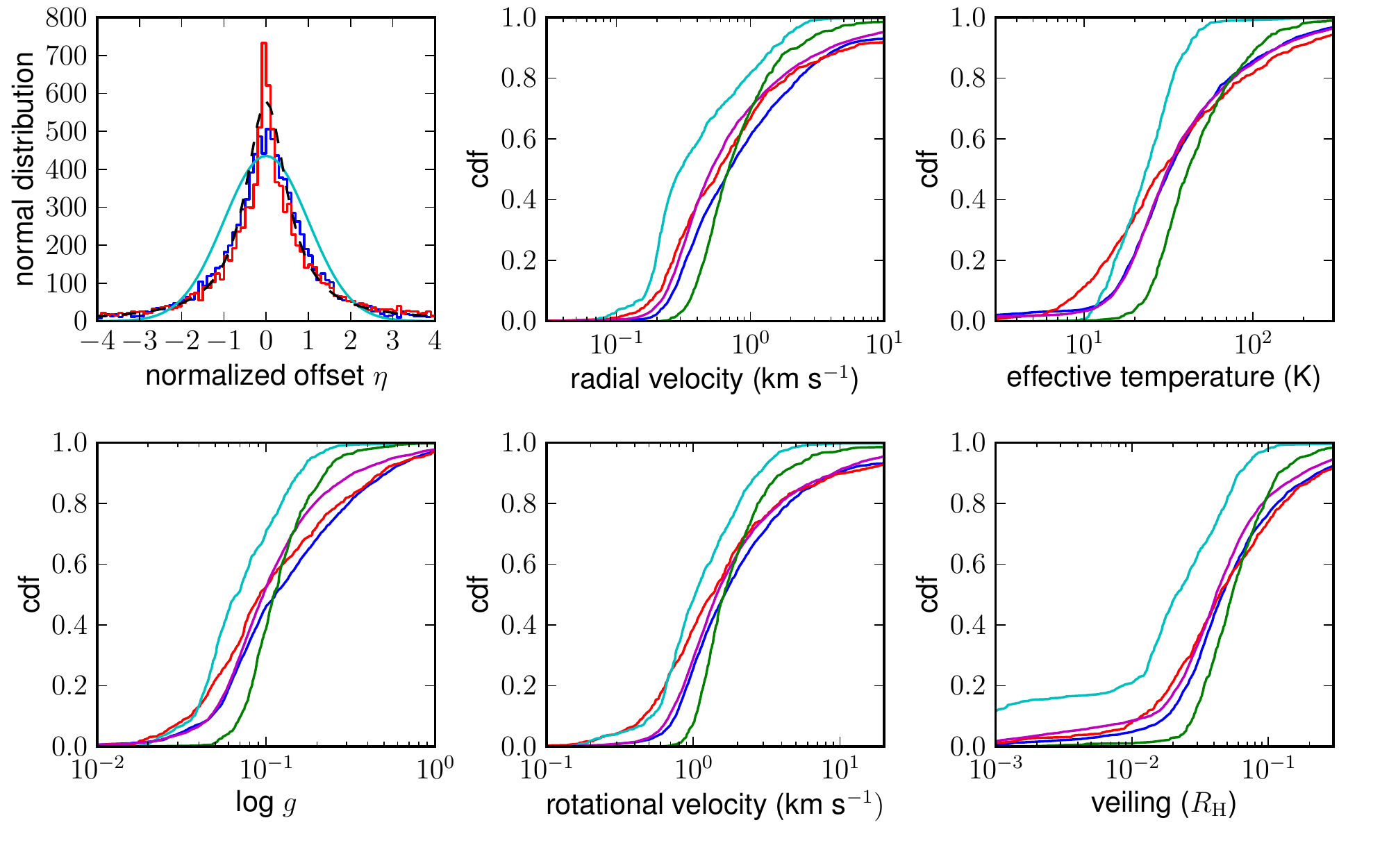}
\caption{\label{fig:dist_unc}Upper-left panel: The distribution of offsets normalized to the uncertainty calibrated to the epoch-to-epoch variability from equation \ref{eq:sigtrue} for the effective temperature (blue) and the radial velocity (red). These distributions are very similar and are better described by a Cauchy-Lorentz distribution with a width of 0.6 (black dashed) than a Gaussian distribution (cyan solid). Other panels: The distribution of the corrected parameter uncertainties for IC 348 (blue), NGC 1333 (red), Orion A (magenta), NGC 2264 (green), and the Pleiades (cyan).}
\end{figure*}

The uncertainty in the stellar parameters is poorly described by a Gaussian distribution. This is illustrated in the upper-left panel of Figure \ref{fig:dist_unc} for both a parameter that should show no intrinsic variation between epochs (i.e., the effective temperature in blue) and a parameter that could for some stars vary between epochs (i.e., the radial velocity in red). Both of these distributions are very poorly described by the Gaussian, but are well described by a Cauchy-Lorentz distribution with a width of 0.6 \(\sigma\), which gives that the probability of the true parameter value being \(p_{\rm true}\) is given by:
\begin{equation}
  P(p_{\rm true}) = \frac{1}{\pi} \left(1 + \frac{(p_{\rm true} - p_{\rm meas})^2}{(0.6 \sigma_{\rm meas})^2}\right)^{-1}, \label{eq:Cauchy}
\end{equation}
where \(p_{\rm meas}\) is the measured parameter value and \(\sigma_{\rm meas}\) is the measurement uncertainty calibrated to the epoch-to-epoch variability (equation \ref{eq:sigtrue}) as quoted in the online tables. The cumulative distribution of these measurement uncertainties in the various clusters have been plotted in the other panels of Figure \ref{fig:dist_unc}. Equation \ref{eq:Cauchy} describes the statistical uncertainties in the stellar parameters due to noise. In the following sections we will look at potential systematic offsets in the stellar parameters.

\subsubsection{Velocity systematics}
\label{sec-3-2-2}
\label{sec:vzero}
To compare our APOGEE derived radial velocities with previous radial velocity observations and measurements of the local gas velocity we need to characterize any systematic offset in our radial velocities. More importantly to measure the velocity dispersion of a cluster is that this systematic offset should not depend on other stellar properties.

\begin{figure}[htb]
\centering
\includegraphics[width=\linewidth]{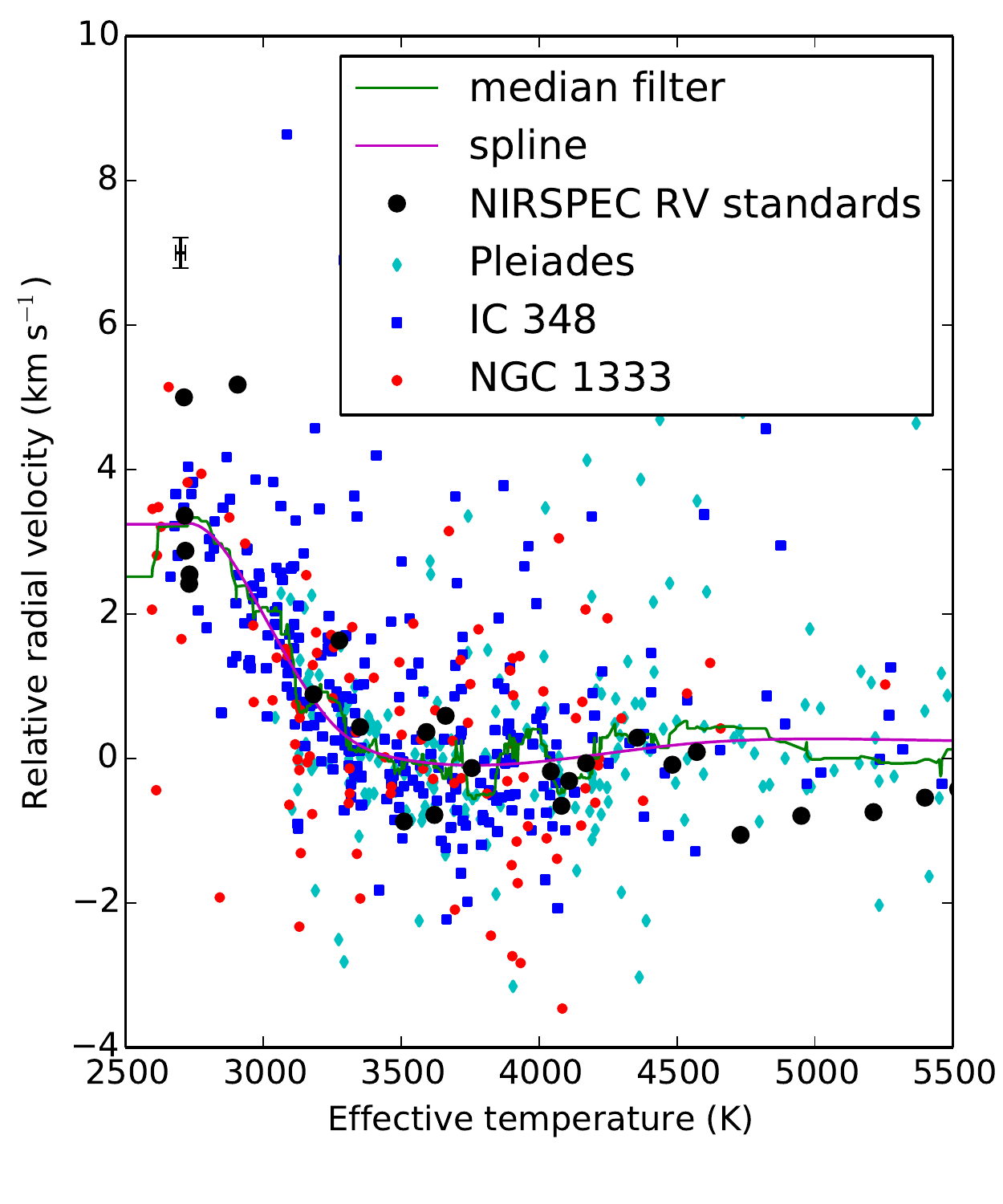}
\caption{\label{fig:vzero_Teff}The trend of the radial velocity with effective temperature. For all stars with a radial velocity precision better than 1 km s\(^{-1}\) in IC 348 (blue squares), the Pleiades (cyan diamonds), and NGC 1333 (red circles) the weighted means of the radial velocities measured over all observable epochs are plotted relative to the median radial velocity for stars between 3500 and 5000 K. The median (1-sigma) statistical uncertainty has been plotted in the upper left. In black the radial velocities derived from applying the IN-SYNC spectral parameter pipeline to NIRSPEC H-band spectra of radial velocity standards from \citet{Prato07} are plotted relative to their literature values. Finally, the green trend line is computed by median-filtering the individual radial velocity offsets and the magenta line shows the best-fit cubic spline through this trend line.}
\end{figure}

Starting by examining the internal consistency of our APOGEE radial velocities, we find a systematic offset of a few km s\(^{-1}\) for the coolest stars (\(T_{\rm eff} < 3500\) K) with respect to the hotter stars in the same cluster. Figure \ref{fig:vzero_Teff} shows this trend for three young clusters observed with APOGEE, namely IC 348 (blue squared), the Pleiades (cyan diamonds), and NGC 1333 (red circles). To show the similarity of this trend in all three clusters, in Figure \ref{fig:vzero_Teff} we plot the stellar radial velocities relative to the systematic velocity of the cluster they are in (which we estimate as the median of the radial velocities of stellar cluster members between 3500 and 5000 K). If this offset were real the low-mass stars with \(T_{\rm eff} < 3000\) K would separate from the cluster containing the hotter stars in less than a few Myr. Because such a translational separation based on mass has never been observed, we conclude that the trend of the mean radial velocity with temperature is unphysical and is caused by a systematic error in the measured radial velocities.

We are able to reproduce this systematic offset with effective temperatures under a wide variety of conditions. The same systematic offset was found in the SYNTHVHELIO radial velocities measured for cool dwarves by the APOGEE radial velocity pipeline. The APOGEE radial velocity pipeline measures SYNTHVHELIO values by cross-correlating, not \(\chi^2\)-minimizing, each APOGEE spectrum against its best-fitting synthetic template from the APOGEE radial velocity mini-grid.  The APOGEE radial velocity mini-grid is primarily drawn from a set of synthetic spectra calculated by the APOGEE/ASPCAP team from a custom line list (Shetrone et al., in prep) and Kurucz model atmospheres for warmer stars (\(T_{\rm eff} >\) 3500 K), but also includes BT-Settl models for \(T_{\rm eff} <\) 3500 K to extend the grid to better cover the coolest dwarfs \citep{Meszaros12}. The presence of a similar systematic offset in the SYNTHVHELIO values measured for the coolest dwarfs in IC 348, NGC 1333, and the Pleiades suggest that the offset is related to the shared usage of the BT-Settl model grid, rather than the distinct and independent radial velocity measurement algorithms incorporated into the main APOGEE and the IN-SYNC specific analysis pipelines.

To rule out instrumental effects as a potential cause for this apparent radial velocity trend with stellar effective temperature, we first measured multiple independent radial velocity values for each APOGEE spectrum, treating portions of the spectrum sampled by the three different detectors independently. These independent radial velocity measurements showed the same trend with effective temperature, indicating that the cause was not isolated to one or two prominent features, but rather affected all parts of the spectra of the coolest stars equally. As a second check on potential instrumental effects that could be causing this trend, we also applied our fitting procedure to NIRSPEC H-band spectra of well-established radial velocity standards obtained and made publicly available by \citet{Prato07}. The radial velocity offsets we measure with respect to these stars' known radial velocities, which were determined using empirical templates, are included in Figure \ref{fig:vzero_Teff}, and show the same trend with effective temperature as the APOGEE spectra, indicating that the radial velocity trend is not an artifact of a particular instrument or reduction procedure. 

Having ruled out potential instrumental and algorithmic causes for this radial velocity trend with effective temperature, we then investigated the model spectra themselves as a potential cause of this effect. Re-fitting the APOGEE spectra with synthetic spectra from the Gaia-ESO grid \citep{Husser13}, we found again the same radial velocity trend with effective temperature, indicating that the cause of this affect is not specific to a single model grid. The systematic RV offset does disappear if we adopt a grid lacking spectra cooler than 3500 K, which indicates that the problem lies with the cooler models. With this trend seemingly related to multiple model grids, most prominent at cool temperatures, and prevalent throughout the full H-band spectrum, we suspect the molecular line-lists as a potential cause of the RV offset, particularly the water lines that become prominent at these late-type stars.

We subtract this systematic offset from each of our radial velocity measurements to provide a stellar radial velocity free of this systematic error. We characterize the offset by fitting a cubic spline to the median-filtered trend of radial velocity with effective temperature considering all three regions (magenta and green lines in Figure \ref{fig:vzero_Teff} respectively). We then use this spline to assign a radial velocity correction to every star based on its effective temperature. No radial velocity correction is applied for stars hotter than 4000 K, where the correction is comparable to the statistical uncertainties. The corrected radial velocities have been included in the online tables.

This correction homogenizes the radial velocity scale at all temperatures to that of the stars hotter than 3500 K, however it does not guarantee that this velocity scale does not have systematic offsets with respect to those of other studies. On the one hand we measure radial velocities systematically \emph{lower} by \(450 \pm 120\) m s\(^{-1}\) than the RV standards from \citet{Prato07} from their NIRSPEC H-band spectra (Figure \ref{fig:vzero_Teff}). On the other hand in the Pleiades we measure radial velocities systematically \emph{higher} by \(400 \pm 60\) m s\(^{-1}\) than the radial velocities measured for the same stars with CORAVEL \citep{Mermilliod09}. The latter offset is consistent at the 2 sigma level with the radial velocity offset of \(602 \pm 116\) m s\(^{-1}\) found between APOGEE and CORAVEL radial velocities for stars in Coma Ber \citep{Terrien14}. We decided not to calibrate our radial velocities to either one  of these systems, because of the different directions of the systematic offsets with respect to the radial-velocity standards from \citet{Prato07} and \citet{Mermilliod09}. From this analysis we estimate a systematic uncertainty in the calibration of our radial velocities on the order of 0.5 km s\(^{-1}\).

\subsubsection{Systematic temperature offsets}
\label{sec-3-2-3}
\label{sec:Teff}
\begin{figure}[htb]
\centering
\includegraphics[width=\linewidth]{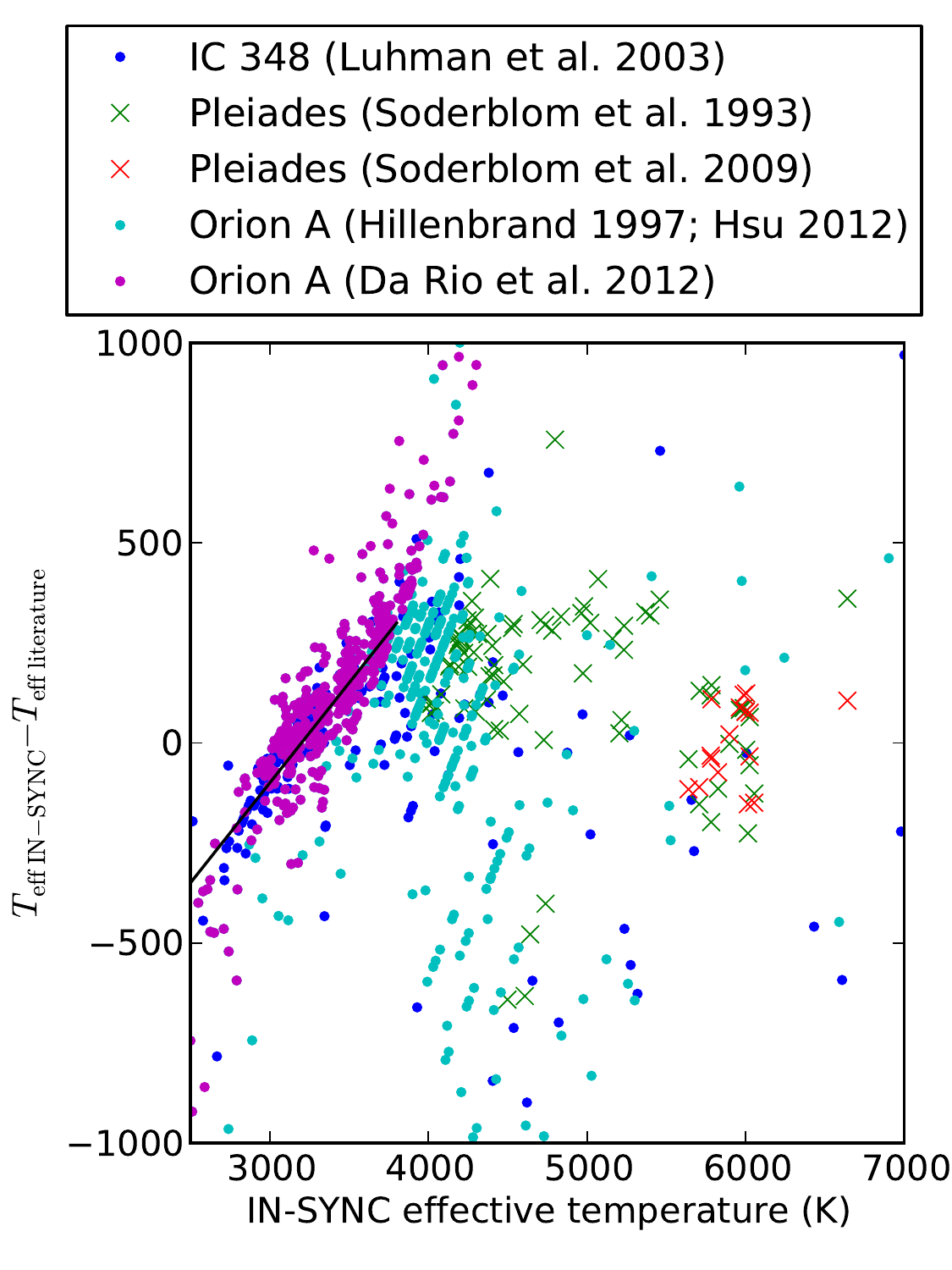}
\caption{\label{fig:Teff_comp}A comparison of the measured IN-SYNC/BT-Settl effective temperatures with literature values for the same stars. The effective temperature of a star observed with APOGEE and a previous survey is plotted as a cross for Pleiades stars and a dot for the younger stars in IC 348 and Orion A. Blue dots compare the effective temperature derived in IC 348 to those derived from both visible and NIR spectral types from a variety of sources collected by \citet{Luhman03}. The green crosses compare to the effective temperature in the Pleiades from visible echelle spectra from the Hamilton spectrograph on the Lick Observatory \citep{Soderblom93} and the red crosses compare with similar data for solar-like stars in the Pleiades from \citet{Soderblom09}. The magenta dots compare to the effective temperatures in Orion A from deep visible photometry with dedicated medium-band filters \citep{Da-Rio12a}. Finally, the cyan dots compare to the effective temperatures in Orion A collected by Nicola Da Rio from spectral types from \citet{Hillenbrand97} and \citet{Hsu12}. The black line illustrates the suggested conversion to the effective temperature scale from \citet{Luhman99} from equation \ref{eq:Luhman}.}
\end{figure}

Here we study the accuracy of the effective temperature derived from the APOGEE spectra. The offset between the \(T_{\rm eff}\) we measure and the \(T_{\rm eff}\) values given for the same stars in several other studies are plotted in Figure \ref{fig:Teff_comp}. The literature \(T_{\rm eff}\) are mostly derived from visible spectroscopy and photometry, although a few are based on near-infrared spectral types. We will discuss the accuracy of the \(T_{\rm eff}\) for stars in three temperature ranges: cool stars with \(T_{\rm eff} < 3800\) K, intermediate stars with 3800 K \(< T_{\rm eff} < 4200\) K and hot stars with \(T_{\rm eff} > 4200\) K.

For the coolest stars ($T_{\rm eff} < 3800$ K) we find a temperature-dependent systematic offset ranging from -300 K to +300 K, when comparing our effective temperatures with the effective temperatures derived from visible and near-infrared spectral types in IC 348 by \citet{Luhman03} and with the effective temperatures from dedicated red medium-band filters sensitive to TiO lines in Orion A \citep{Da-Rio12a} (Figure \ref{fig:Teff_comp}). The strong agreement in the systematic offset in the effective temperatures relative to both \citet{Luhman03} and \citet{Da-Rio12a} implies that the effective temperature scale obtained from the APOGEE observations in the H-band using the BT-Settl grid disagrees with the conversion from spectral type to effective temperature from \citet{Luhman99}. This conversion was both used in \citet[][except for a small change for $T_{\rm eff} < 2700$ K]{Luhman03} and to calibrate the photometry to effective temperatures in \citet{Da-Rio12a}. The spectral types themselves are an unlikely origin of the systematic offset as they come from a wide variety of literature sources and have been derived from both visible and near-infrared spectra. The IN-SYNC effective temperatures can be converted to the effective temperature scale from \citet{Luhman99} by:
\begin{equation}
T_{\rm Luhman\ (1999)} = T_{\rm IN-SYNC} - 0.5 (T_{\rm IN-SYNC} - 3200 {\rm K}), \label{eq:Luhman}
\end{equation}
where \(T_{\rm IN-SYNC}\) is the measured effective temperature from the online tables and \(T_{\rm Luhman\ (1999)}\) is according to the temperature scale from \citet{Luhman99}. Ignoring the systematic offset a scatter remains of only \(\sim 80 K\) over this temperature range, which is in line with the precision of half of a spectral subtype claimed by \citet{Luhman03} and \citet{Da-Rio12} and is larger than our estimated uncertainties of \(\sim\) 20 K (Figure \ref{fig:dist_unc}). A likely cause of this remaining scatter is the tendency of the best-fit effective temperatures to cluster around every other BT-Settl grid point in effective temperature (Figure \ref{fig:Teff_logg}).

For hotter stars (\(T_{\rm eff} > 3800 K\)) the scatter in the comparison between IN-SYNC and literature effective temperature increases. Up to effective temperatures of about 4200 K, we find that for the youngest stars (i.e., the blue dots for IC 348 stars and cyan dots for Orion A stars) we overestimate the effective temperature by about 200 K with respect to literature values. However, there is a significant population of stars with IN-SYNC temperatures in this range (3800 to 4200 K) with much higher literature effective temperature. For even high IN-SYNC effective temperatures (hotter than 4200 K) the scatter for these young stars becomes on the order of 500 K.

For the older stars in the Pleiades (marked by crossed in Figure \ref{fig:Teff_comp}) the literature temperatures agree much better with the IN-SYNC temperatures among these early-type stars with typical offsets on the order of only about 100 K. This suggests that this large scatter is limited to the early-type (i.e., hotter than roughly 4000 K), young stars. We speculate that this might be caused by the large magnetic fields in these young stars. The Zeeman broadening from these magnetic fields has not been included in our model of the APOGEE spectra and could potentially cause significant biases.
\subsubsection{Accuracy in surface gravities}
\label{sec-3-2-4}
\label{sec:logg}
\begin{figure*}
\centering
\includegraphics[width=1\textwidth]{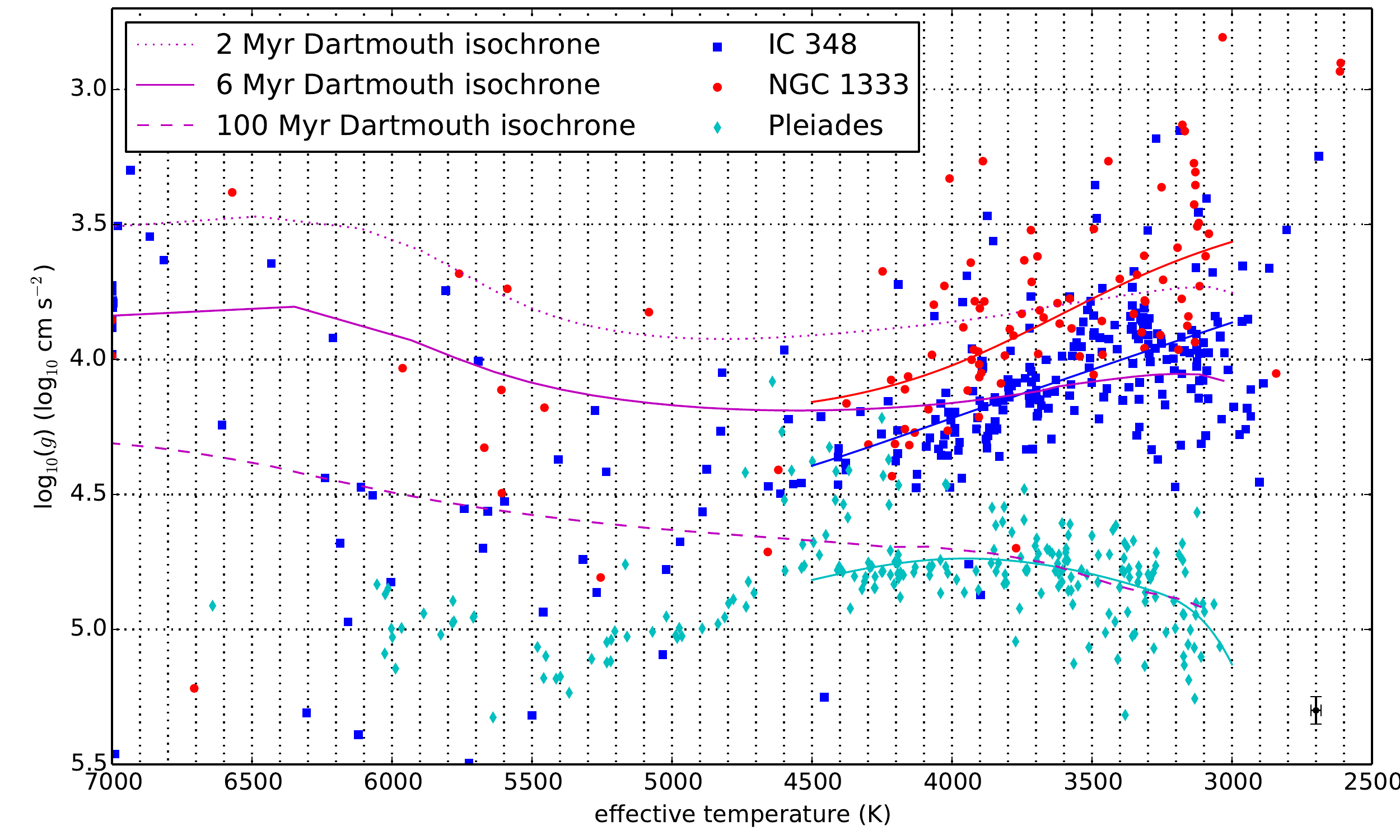}
\caption{\label{fig:Teff_logg}The distribution of spectroscopic effective temperatures and surface gravities (blue squares for IC 348; red circles for NGC 1333, and cyan diamonds for the Pleiades) with trend lines overplotted in the same color between 3000 and 4500 K. Only stars with \(\sigma_{\rm Teff} < 100\) and \(\sigma_{\log(g)} < 0.1\) have been included for clarity. The median (1-sigma) statistical uncertainty of the remaining stars is shown in the lower right. Overplotted in magenta are the Dartmouth isochrones \citep{Dotter08} at 2 Myr (dotted), 6 Myr (solid), and 130 Myr (dashed). The BT-Settl grid contains a model spectrum at every intersection of the black dotted lines and extends to lower effective temperatures and lower surface gravities than plotted here.}
\end{figure*}

We also determine the accuracy of the measured surface gravities by comparing the observed effective temperature vs. surface gravity diagrams for IC 348, NGC 1333, and the Pleiades with those expected from isochrones (Figure \ref{fig:Teff_logg}). For the cooler temperatures (\(T_{\rm eff} < 4500\) K) the stars in the different clusters clearly differ in their surface gravity with the stars in NGC 1333 having the lowest surface gravities (and hence youngest contraction ages) and the stars in the Pleiades having the highest surface gravities. We will show in Section \ref{sec:lum_spread} that the surface gravities are even precise enough to detect a spread in stellar radii within IC 348. The surface gravities in this temperature range agree reasonably well with the Dartmouth isochrones expected at their respective ages, namely \(\sim 1-2\) Myr for NGC 1333, \(\sim 3-6\) Myr for IC 348 \citep{Luhman03, Bell13a}, and \(\sim 130\) Myr for the Pleiades \citep{Stauffer98, Barrado-y-Navascues04}. 

The increased spread and offset from the isochrones in surface gravities for hotter stars (\(T_{\rm eff} > 4500\) K) suggest a lower precision and accuracy for the hotter stars, although the older pre-main sequence stars in the Pleiades retain a higher surface gravity than those in the younger clusters. These surface gravities in the Pleiades are almost certainly overestimated, because they would imply that the stars in the Pleiades are significantly smaller than main sequence stars. For example, the Sun has a \(\log g\) of \(\sim 4.4\) at its effective temperature of \(\sim 5800\) K, which would imply that the solar radius is about twice as large as the radius of the stars with similar temperature in the Pleiades, which have \(\log g \approx 5\) (assuming that the Pleiades stars at that temperature are about 1 solar mass). For the younger stars in IC 348 and NGC 1333 the scatter in surface gravity is very large in this temperature range (\(T_{\rm eff} > 4500\) K), which is reminiscent of the large scatter in the comparison of effective temperatures with literature values that we found for these hot, young stars. This implies that the effective temperatures and surface gravities of stars hotter than roughly 4500 K should be treated with great caution, especially in the younger clusters and even though these stellar parameters seem to be more precisely measured in the Pleiades the surface gravity for these older stars is probably overestimated. However, most stars observed in our sample are significantly cooler and for these stars the effective temperature and surface gravity appear to be measured to high precision (with perhaps some systematic offsets).

\subsubsection{Accuracy of the rotational velocities}
\label{sec-3-2-5}
\label{sec:vsini}
The stellar rotation can be constrained either through the rotational broadening of stellar spectra or the measurement of the rotation period from photometric variability, caused by starspots that rotate with the stellar surface. We will use both of these measurements to validate our rotational velocities.

\begin{figure}[htb]
\centering
\includegraphics[width=1\columnwidth]{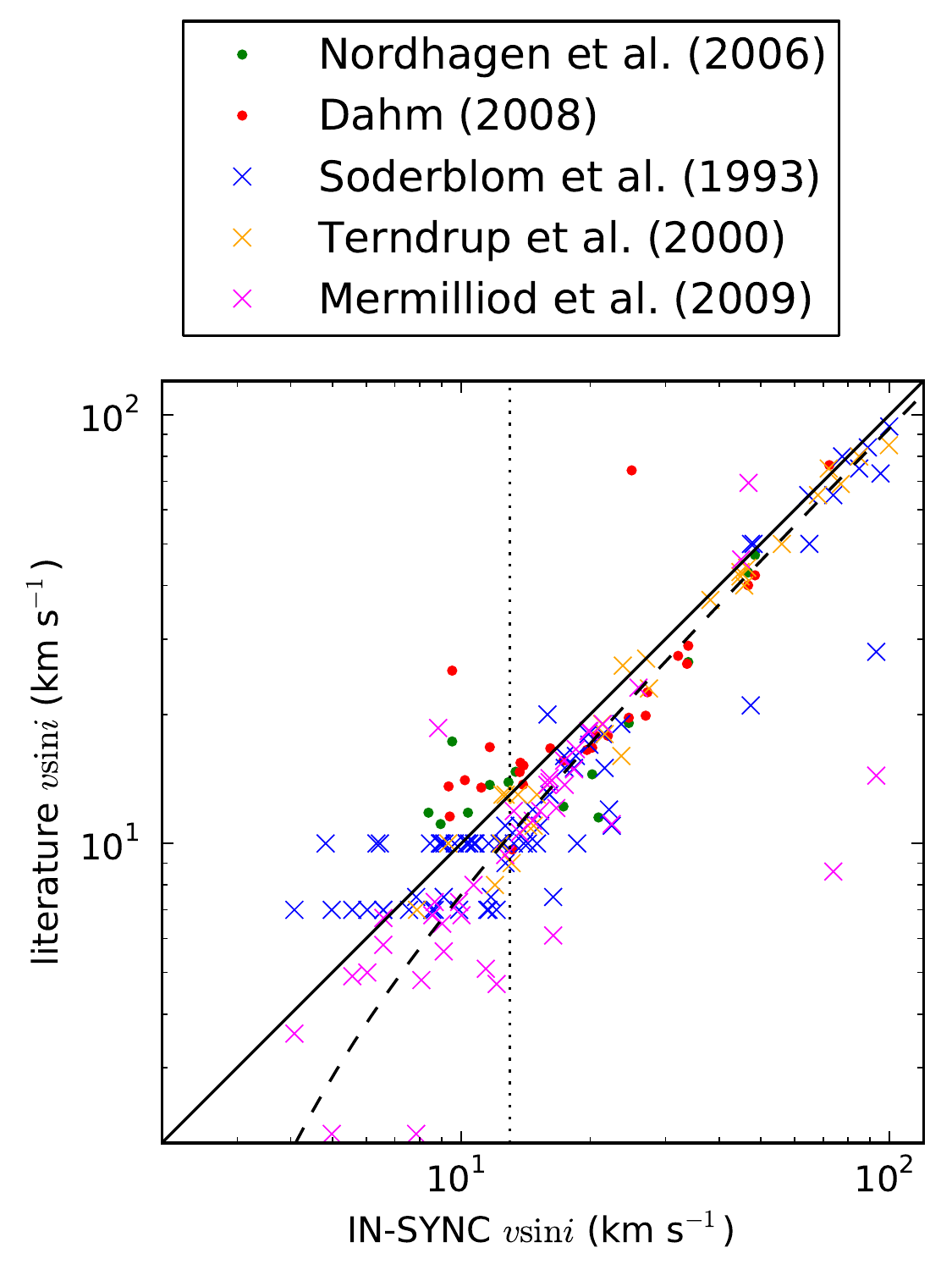}
\caption{\label{fig:vsini_comp}Rotational velocities measured from the APOGEE spectra compared with literature values in IC 348 (circles) and the Pleiades (crosses). The solid line shows the one-to-one correspondence, while the dashed line shows the expected trend for a relative overestimate of the IN-SYNC \(v \sin i\) of 5\% as well as an absolute offset of 2 km s\(^{-1}\) (equation \ref{eq:vsini_true}). The broadening due to rotation matches the spectral resolution at the vertical dotted line at \(v \sin i = 13\) km s\(^{-1}\). The typical epoch-to-epoch variability is around 1 km s\(^{-1}\), much smaller than the observed \(v \sin i\) except for the slowest rotators.}
\end{figure}

Figure \ref{fig:vsini_comp} compares the rotational velocities measured from the APOGEE spectra in this work with those previously measured from high-resolution visible spectra in the Pleiades \citep{Soderblom93, Terndrup00, Mermilliod09} and IC 348 \citep{Nordhagen06, Dahm08}. The APOGEE rotational velocities tend to be systematically higher than the literature values. When fitting this offset by eye we find two separate contributions. The first is a relative offset in the \(v \sin i\) of a few percent, which is most obvious among the rapid rotators in the comparison with the rotational velocities from \citet{Soderblom93} and \citet{Dahm08}. This offset might be caused by the lower limb darkening in the near-infrared compared to the visible, which leads to a larger spectral broadening for the same \(v \sin i\) \citep[e.g.,][]{Magic14}. In addition to this relative offset there is an absolute offset in the \(v \sin i\) of about 2 km s\(^{-1}\), which causes the literature and IN-SYNC \(v \sin i\) to diverge for the slow rotators in the logarithmic scale of Figure \ref{fig:vsini_comp}. Only in the comparison with the rotational velocities from \citet{Mermilliod09} does this trend continue to the slowest rotators, which probably reflects that the resolution limit of the other studies was reached around 10 km s\(^{-1}\). This offset is small compared to the width of the spectral resolution (APOGEE's \(R \sim 22,500\) corresponds to 13 km s\(^{-1}\)) and is likely caused by uncertainties in the instrumental line profile. Indeed the actual line profile of APOGEE is not given by a Gaussian with a single resolution, but depends on both pixel and fiber. A more accurate treatment of this line profile is needed to reliably measure the low rotational velocities. When we combine this relative offset and absolute offset we find:
\begin{equation}
v \sin i_{\rm true} = 0.95 (v \sin i_{\rm IN-SYNC} - 2 \text{km s}^{-1}). \label{eq:vsini_true}
\end{equation}
By labeling the resulting \(v \sin i\) as true, we imply that the literature rotational velocities are probably more accurate than those derived here from the APOGEE spectra. Next we will show that this is indeed the case by comparing the IN-SYNC \(v \sin i\) with rotational periods.

\begin{figure}[htb]
\centering
\includegraphics[width=1\columnwidth]{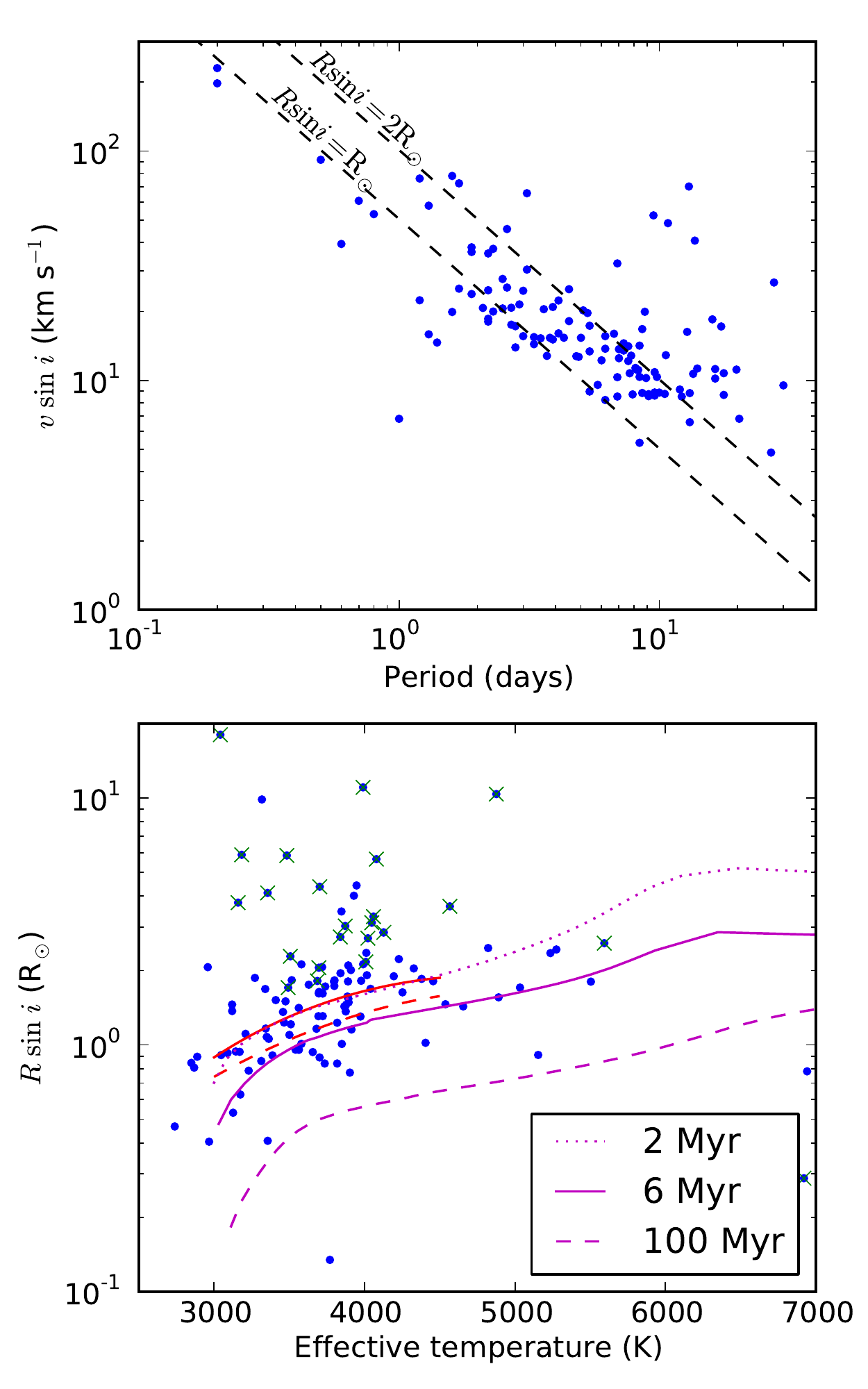}
\caption{\label{fig:vsini_dist}Upper panel: Rotational velocity versus rotational period from \citet{Cieza06}. Faster rotators have shorter periods as expected. The expected correlation for \(R \sin i = {\rm R_{\odot}}\) and \(R \sin i = 2 {\rm R_{\odot}}\) have been overplotted as dashed lines. Lower panel: Stellar radii from the multiplication of the rotational velocities and periods versus effective temperature. The red solid line shows the trend line (obtained from fitting a spline to median-filtered \(R \sin i\)). The red dashed line shows the trend, when the \(v \sin i\) from equation \ref{eq:vsini_true} is used to compute the \(R \sin i\). Slow rotators with periods larger than 10 days have been marked by green crosses and have not been taken into account for computing the trend line. The magenta lines in this panel are the 2, 6, and 130 Myr isochrones.}
\end{figure}

\citet{Cieza06} present 143 rotational periods derived from photometric variability in IC 348, of which 75 were measured for the first time by these authors, while the others were collected from \citet{Cohen04}, \citet{Littlefair05}, and \citet{Kiziloglu05}. Out of these 143 stars with rotational periods 120 have APOGEE spectra. For a fixed radius \(R\) the rotational periods should be negatively correlated with the rotational velocities, which we indeed find in this dataset (see upper panel of Figure \ref{fig:vsini_dist}). Every rotational period a parcel of gas at the equatorial surface of the star will have to travel a total distance of \(2 \pi R\), which implies that we can calculate the radius of the star from the product of the rotational velocity and period. Because we can only measure the projected rotational velocity \(v \sin i\) we can not compute the radius directly, but only \(R \sin i\) through
\begin{equation}
  R \sin i = \frac{P \cdot v \sin i}{2 \pi}. \label{eq:rsini}
\end{equation}
The upper panel of Figure \ref{fig:vsini_dist} shows that the IN-SYNC rotational velocity and the \citet{Cieza06} rotational periods are negatively correlated as expected for a roughly constant stellar radius. However the slower rotators do seem to cluster a bit more around the line of \(R \sin i = 2 R_\odot\) rather than the \(R \sin i = R_{\odot}\), where we find more of the rapid rotators. This could simply reflect that more rapid rotators are somewhat smaller in size, however as we will show in Section \ref{sec:age_spread} in IC 348 more rapid rotators are actually larger in size. This is hence further evidence that there is indeed an absolute overestimate in the IN-SYNC rotational velocities, which can explain the somewhat higher \(R \sin i\) found for slow rotators.

The lower panel in Figure \ref{fig:vsini_dist} shows the computed \(R \sin i\) distribution, compared with the radii from the Dartmouth isochrones. These \(R \sin i\) are surprisingly large, with the trend line (red solid line based on median filtering) centered on the 2 Myr isochrone, rather than the 6 Myr isochrones such as the surface gravities in Figure \ref{fig:Teff_logg} or the luminosities, which we will discuss in Section \ref{sec:lum} (Figure \ref{fig:absJ}). This once again suggests a systematic overestimate in the \(v \sin i\), which is much more likely than a systematic offset in the periods, because the periods are typically measured over baselines that are much longer than the rotation periods (leading to a high precision) and the periods appear to be stable over multiple rotations \citep{Nordhagen06}. When we use the corrected \(v \sin i\) from equation \ref{eq:vsini_true} to recompute the trend of \(R \sin i\) we find a much lower typical \(R \sin i\) (shown by the red dashed line) more consistent with the evolutionary stages estimated from the stellar surface gravities and luminosities.

So we conclude that the IN-SYNC rotational velocities are systematically overestimated by a relative offset of 5\% (probably due to the adopted profile for spectral broadening) and by an absolute offset of about 2 km s\(^{-1}\) (probably due to uncertainties in the instrumental line profile). Evidence for this overestimation is found from both the comparison with literature rotational velocities in the Pleiades and IC 348 as well as the large \(R \sin i\) found for a subset of stars in IC 348 with rotational periods. The absolute offset of about 2 km s\(^{-1}\) in this overestimate is also separately confirmed by the larger \(R \sin i\) found for slower rotators with respect to rapid rotators, which is the opposite of the actual physical trend in IC 348.

\section{Extinction and stellar luminosities}
\label{sec-4}
\label{sec:phot}
\subsection{Extinction}
\label{sec-4-1}
\label{sec:Av}
We derive the extinctions to the observed young pre-main sequence stars from the color excess in J-H with respect to the typical J-H of stars in the Pleiades with the same effective temperature. Although the Pleiades with an age of \(\sim 130\) Myr, is significantly older than the 1 - 6 Myr old clusters targeted in this survey, the cluster serves as an excellent calibrator, because:
\begin{enumerate}
\item The Pleiades has also been observed by APOGEE, so that we can analyze the spectra of 216 Pleiades members in a manner consistent with our IN-SYNC targets and place them all on the same APOGEE-based \(T_{\rm eff}\) scale.
\item For a subset of stars in IC 348 and the Pleiades we have visible g, r, i, z photometry from \citet{Bell12a, Bell13a}. In both clusters this photometry has been taken from the Wide Field Camera on the Isaac Newton Telescope (INT-WFC) and has been reduced by the same team, which should minimize any systematic offsets.
\item A similar uniform photometric dataset for all stars is available in the near-infrared J, H, and Ks bands from the 2MASS survey \citep{Skrutskie06}.
\item Only a small extinction of E(B-V) = 0.044 is present in the Pleiades \citep[e.g.,][]{Breger86} with little differential extinction\footnote{The differential extinction in the Pleiades (\(A_{\rm V} \sim 1.3\)) caused by a CO cloud covering the southwest region of the cluster \citep{Breger87} is not taken into account in this analysis, because it only covers a very small part of the cluster.}.
\end{enumerate}
\citet{Bell12a} provide a single-star photometric locus for their visible and the 2MASS infrared photometry in the Pleiades. For every star in the Pleiades we locate the point on the locus with the most similar visible and near-infrared photometry by minimizing the unweighted residuals between the star's photometry and that of the locus point in question (hence ignoring the small differential extinction in the Pleiades). Only Pleiades stars with both visible and near-infrared photometry are considered. These Pleiades stars are then used to calibrate the relationship between the spectroscopic effective temperature scale in this work and the grizJHKs colors from the Bell et al. single-star photometric locus.

\begin{figure}[htb]
\centering
\includegraphics[width=\linewidth]{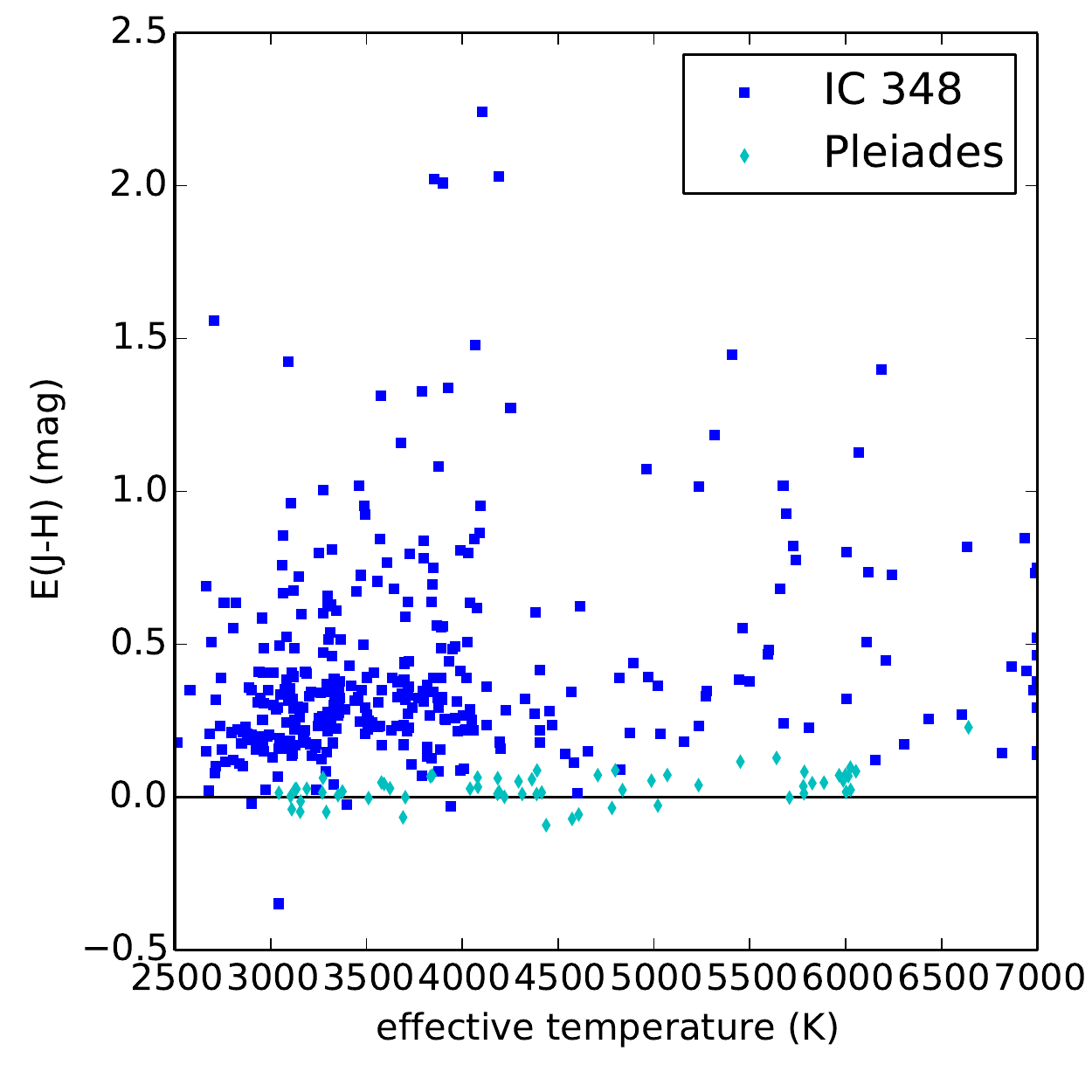}
\caption{\label{fig:EJH}The color excess in J-H for members of IC 348 (blue dots) and the Pleiades (cyan diamonds), calculated with respect to the Pleiades single star sequence and shown plotted against the effective temperature.}
\end{figure}

This single-star locus, calibrated to our effective temperature scale, is then used to covert the effective temperature into an intrinsic color for every star in the targeted young clusters, where we ignore any effect from the age difference between these clusters and the Pleiades. Each star's extinction is then estimated from its 2MASS near-infrared photometry, which is the only photometry available for all our target stars. Specifically the extinction is estimated from the excess in the J-H color (see Figure \ref{fig:EJH}). An additional reddening of E(J-H) = 0.013 is added to correct for the mean extinction in the Pleiades. Here we used \(\frac{\rm E(J-H)}{\rm E(B-V)} = 0.33\) from \citet{Rieke85}.

\begin{figure}[htb]
\centering
\includegraphics[width=\linewidth]{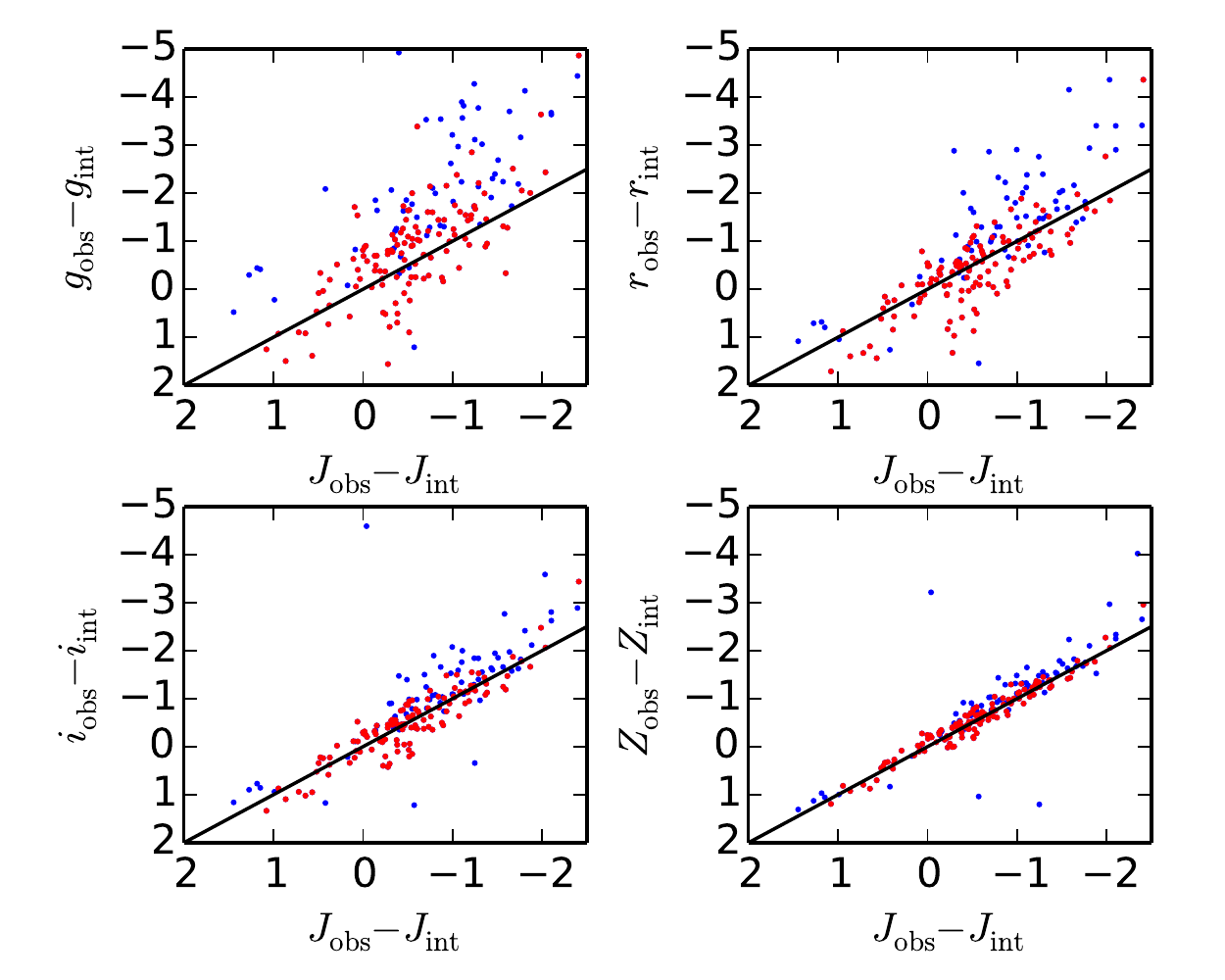}
\caption{\label{fig:mag_excess}The excess magnitudes in various bands in IC 348 with respect to the \citet{Bell12a} single-star locus at the same temperature in the Pleiades after extinction correction. We only include stars with the INT-WFC visible photometry in griZ from \citet{Bell13a}. The black lines in each panel show perfect one-to-one agreement between the excess in the J-band and in the visible. The excess J-band magnitude with respect to the Pleiades correlates very well with the excesses in the visible bands, implying that the excess magnitude is color-independent. The stars have been separated by color into those stars with \(A_{\rm J} > 1\) (blue) and \(A_{\rm J} < 1\) (red).}
\end{figure}

We test the accuracy of this procedure by considering the subset of spectroscopically observed stars in IC 348 that have visible photometry from \citet{Bell13a}. We first correct the extinction in all seven visible and near-infrared bands based on the extinction estimate from the E(J-H). For this extinction correction we use the extinction law for \(R_{\rm V} = 3.1\) from \citet{Cardelli89} integrated over the INT-WFC filters from \citet{Bell12a} and the 2MASS filters from \citet{Cohen03}. We then compute the excess of these extinction-corrected magnitudes with respect to the temperature-calibrated single-star photometric locus. This excess magnitude tells us how much brighter the stars in IC 348 are in apparent magnitude after applying the E(J-H) extinction-correction with respect to the stars with the same effective temperature in the Pleiades. The excesses in the visible and the J-band typically agree within several hundredths to tenths of a magnitude at least for stars with \(A_{\rm J} < 1\) (red in Figure \ref{fig:mag_excess}). This implies that the near-infrared J-H color excess does an excellent job of correcting the (larger) extinction in the visual. Because the visible photometric excesses provide an independent measure of the extinction (for a subset of all stars), we can use the scatter around the \(\Delta i = \Delta J\) line in Figure \ref{fig:mag_excess} to put an upper limit on the uncertainty in the extinction of 0.16 mag in \(A_{\rm J}\). 

The upper panels of Figure \ref{fig:mag_excess} illustrate that there appears to be a subpopulation of stars, that (i) have high extinctions (blue dots), (ii) are relatively bright compared to the Pleiades stars (i.e., towards the right side in the panels), and (iii) show a blue excess (i.e., lie above the black solid line). The larger brightness of these stars might be spurious, because of uncertainties in the large extinctions or because of the larger effect of the adopted extinction law for these stars with higher extinction. However, the detection of a blue excess for these high-extinction stars is robust, which suggests that these stars are accreting. That these blue excesses are found preferentially among stars with higher E(J-H) could be explained if the higher E(J-H) was caused by extinction from the disc or envelope surrounding these actively accreting stars, or by near-infrared emission from the inner part of a circumstellar disc.
\subsection{Stellar luminosities}
\label{sec-4-2}
\label{sec:lum}
\begin{figure}[htb]
\centering
\includegraphics[width=\linewidth]{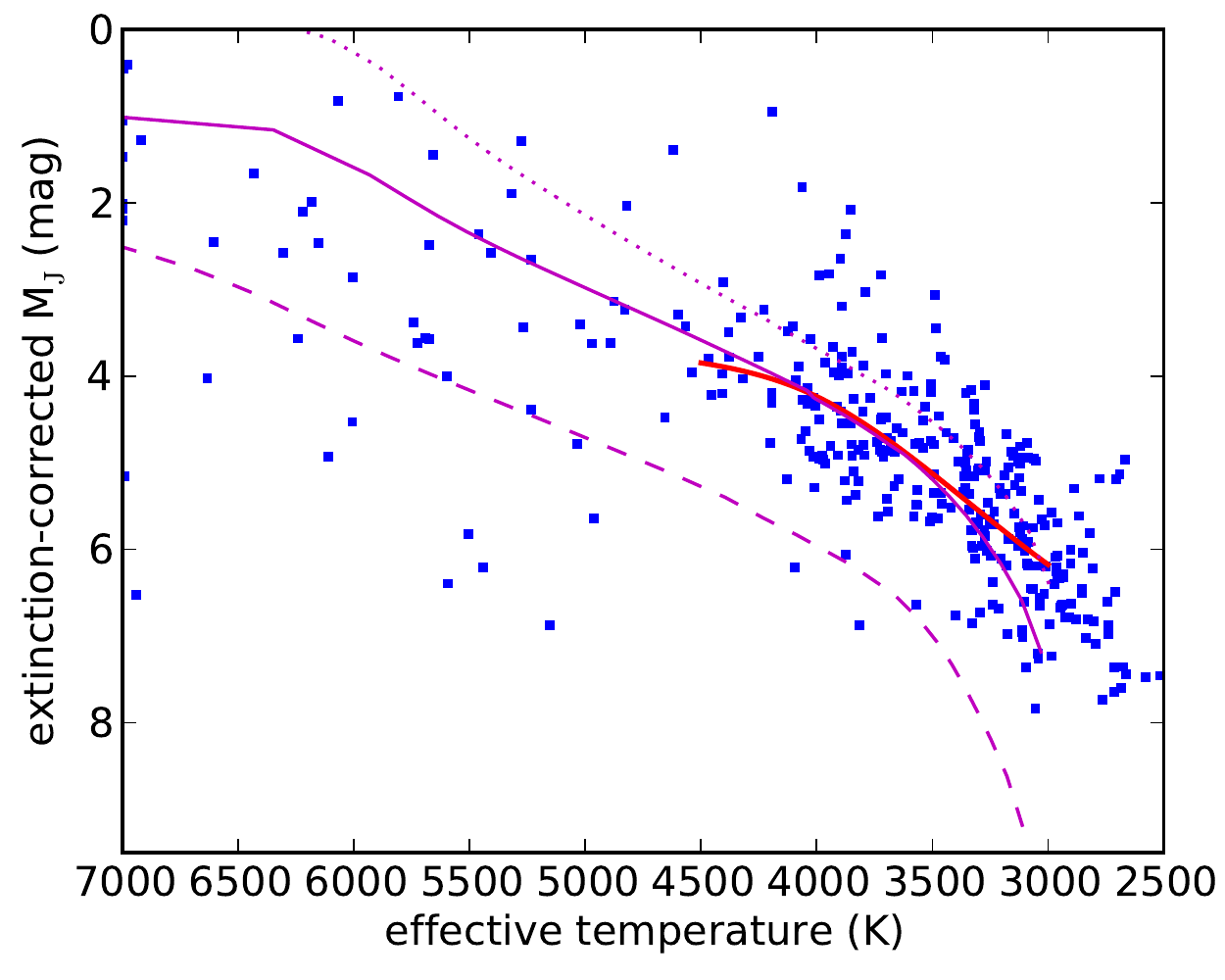}
\caption{\label{fig:absJ}The absolute J-band magnitude after extinction-correction as a function of spectroscopic effective temperature in IC 348, assuming a distance modulus of 6.98. Overplotted are the same Dartmouth isochrones as in Figure \ref{fig:Teff_logg} with the 2MASS J-band magnitude derived from PHOENIX synthetic fluxes \citep{Dotter08}. From top to bottom the ages of the isochrones are 2 Myr (dotted), 6 Myr (solid), and 130 Myr (dashed). In Cottaar et al. (submitted) we will show that most and possibly all stars below the 130 Myr isochrone are most likely field stars. Also shown in red is a spline fit to the median filtered trend of J as a function of \(T_{\rm eff}\) for the IC 348 sample.}
\end{figure}

After extinction-correction and adopting a distance modulus of 6.98 to IC 348 \citep{Ripepi14}, we can derive the absolute magnitudes of the stars in IC 348. Figure \ref{fig:absJ} shows a J-band magnitude-temperature diagram. We overplot the Dartmouth isochrones \citep{Dotter08}, which follow the observed sequence to a much better degree than for either the surface gravities (Figure \ref{fig:Teff_logg}) or the \(R \sin i\) (Figure \ref{fig:vsini_dist} bottom panel). 

\section{Spread in stellar radii in IC 348}
\label{sec-5}
\label{sec:lum_spread}
\subsection{Evidence for a spread in stellar radii}
\label{sec-5-1}
\label{sec:radius_spread}
Previously, \citet{Herbig98}, \citet{Luhman03} and \citet{Muench07} pointed out that at any given effective temperature the stars in IC 348 show a spread in extinction-corrected apparent brightness. If this corresponds to an intrinsic spread in luminosities and hence stellar radii in IC 348, this suggests that the stars have a non-zero range of contraction ages, although other effects can also cause an intrinsic spread in stellar radii (see Section \ref{sec:age_spread}). However, many physical and observational causes can broaden the distribution of extinction-corrected brightness, not only an intrinsic spread in luminosities \citep{Hartmann01, Reggiani11a}. 

Here we use three (nearly) independent estimates of the stellar radii to show that there is an intrinsic stellar radius spread in IC 348, corresponding to an average spread in stellar radius of about 25\% at a single effective temperature. These stellar radius estimators are the extinction-corrected M\(_{\rm J}\) (Section \ref{sec:lum}), the spectroscopic surface gravity \(\log(g)\) (Section \ref{sec:logg}), and the \(R \sin i\) from equation \ref{eq:rsini} (Section \ref{sec:vsini}). To find a spread in the radius estimators at an effective temperature, we need to measure the offset between each star's radius indicator and the median value at that effective temperature. To measure these offsets we subtract from these three stellar parameters the empirical trend lines as illustrated as the blue line in Figure \ref{fig:Teff_logg} and the red lines in Figures \ref{fig:vsini_dist} and \ref{fig:absJ}. The advantage of using empirical trend lines rather than model isochrones is that the result is model-independent and it is not influenced by temperature-dependent systematic offsets in the observed stellar parameters. The main disadvantage is that we can only quote relative radii (or ages) and we can only include stars over the temperature range where the trend with effective temperature can be accurately determined (i.e. between 3000 and 4500 K). Even in this range stars with a temperature uncertainty larger than 100 K are excluded. We also exclude any stars with uncertainties in the surface gravity larger than 0.1 dex and slow rotators (with \(v \sin i < 15\) km s\(^{-1}\)) for which the \(R \sin i\) estimate depends strongly on the resolution (see Section \ref{sec:vsini}). These cuts in the surface gravity and \(R \sin i\) precision are only made for the plots and statistical analyses, where these parameters are actually used.

Even if there is no intrinsic stellar radius spread, we still expect scatter due to noise and other effects that can mimic a stellar radius spread. Here we will discuss for every stellar radius indicator the dominant sources that mimic a stellar radius spread, which we can then compare to the observed spread. For the \(R \sin i\) distribution we include in this spread estimate the effects from the uncertainties in the \(v \sin i\) (which is much larger than the uncertainty in the period) as well as the effect of the inclination given a random orientation\footnote{This is an overestimate of the actual effect of the \(\sin i\), because stars are much more likely to show periodic photometric variability if they are observed at high inclinations (high \(\sin i\)) than low inclinations. This biases the sample of stars with observed periods towards high \(\sin i\).}. For the surface gravity only the measurement uncertainties in this parameter have been included. Finally for the extinction-corrected J a stellar radius spread could be mimicked by the uncertainty of the extinction (which, with an uncertainty of 0.16 mag, dominates the photometric uncertainty in the J-band), as well as the effect of binarity, which we model assuming a flat mass ratio distribution and a J-band flux going as \(F_{\rm J} \approx M^{3}\), which is appropriate for hot companions (with \(T_{\rm eff} > 3500\) K), but slightly underestimates the flux contribution of cooler companions according to the 6 Myr isochrone from \citet{Dotter08}. Conservatively we assume that all stars have an unresolved binary companion. In reality this binary fraction and hence the broadening effect of binary companions on the extinction-corrected J distribution is probably smaller. 

\begin{figure*}
\centering
\includegraphics[width=1\textwidth]{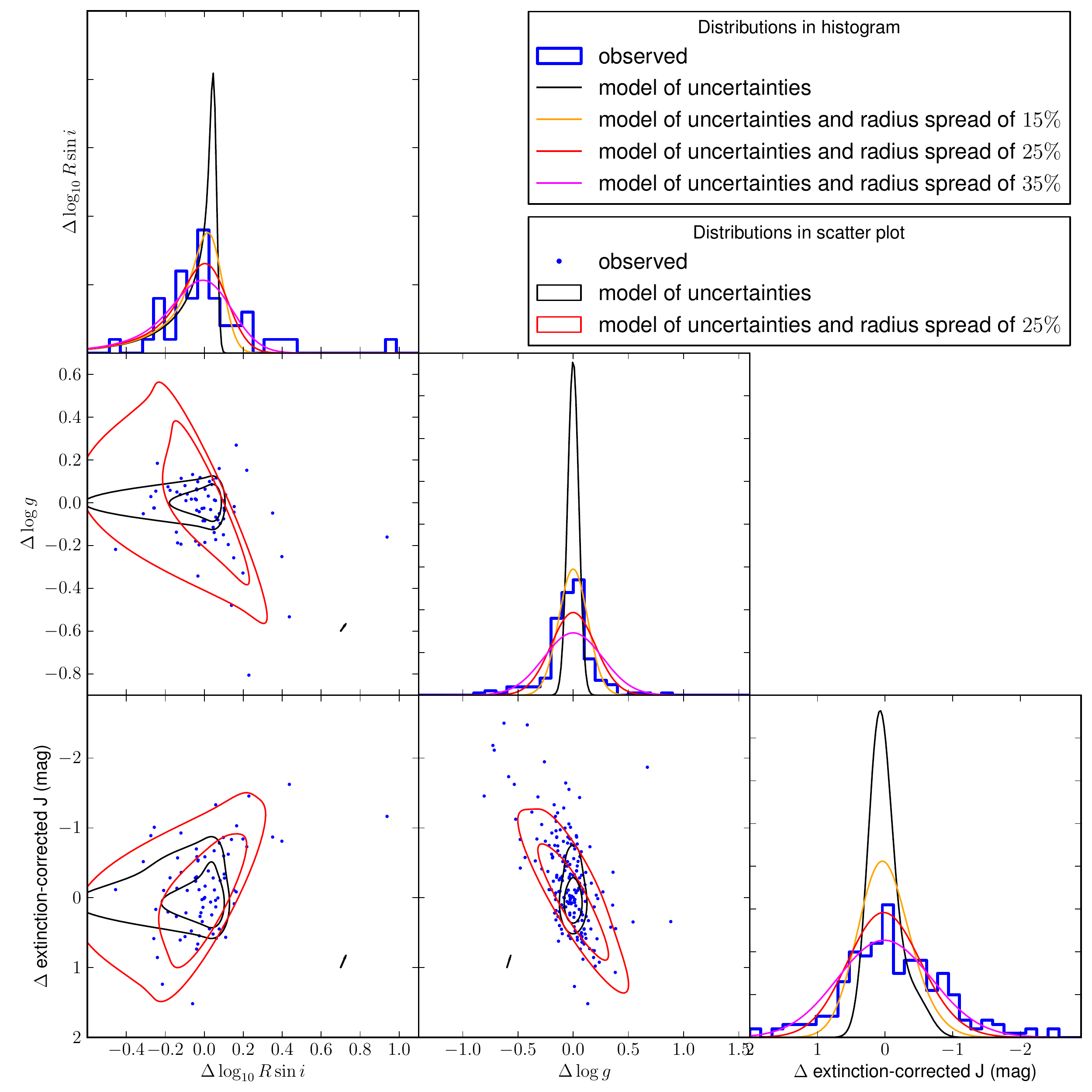}
\caption{\label{fig:triangle}The distributions of three radius estimates, namely the surface gravity (\(\log g\)), extinction-corrected J-band magnitude, and rotational radius (\(\log_{\rm 10} R \sin i\)). All three parameters have been corrected for their dependence on effective temperature by subtracting the trend line with effective temperature. The blue histograms and dots show the distribution of observed stellar parameters. The black distributions illustrate the expected distributions taking into account all effects that could mimic a stellar radius distributions (i.e., measurement uncertainties, binarity, and the projection effect in \(R \sin i\)). The orange, red, and magenta distributions represent the expected distribution given all effects that can mimic a stellar radius distribution as well as a Gaussian intrinsic stellar radius distribution with a width of 15, 25, and 35\% around the median stellar radius at that effective temperature. In the scatter plots the same distributions are plotted by showing the 1 and 2 \(\sigma\) confidence levels, illustrating the expected large correlation between the three radius estimators if there is an intrinsic spread in stellar radii (red distribution) compared with the uncorrelated distribution expected without an intrinsic spread in stellar radii (black distribution).}
\end{figure*}

The histograms in Figure \ref{fig:triangle} compare the observed distribution of these stellar radius indicators (in blue) with the expected distribution if there were no intrinsic stellar radius spread taking into account all sources of an apparent radius spread discussed in the paragraph above (in black). The expected distributions for several Gaussian intrinsic stellar radius spreads have also been included. For all three stellar parameters the width of the observed distributions (blue histogram) is much broader than given by these measurement uncertainties alone (i.e., black distribution), strongly suggesting that there is a finite spread in the stellar radius distribution. Indeed we find that for all three stellar parameters the main peak of the distribution is well modeled by a Gaussian distribution of stellar radii with a width of roughly 25\% around the median radius at that effective temperature (red distribution in Figure \ref{fig:triangle}), although there are too many outliers for this to be formally a good fit. 

Our main evidence for an intrinsic stellar radius spread in IC 348 comes from considering the correlations between the three stellar radius estimators. All of the sources of an apparent stellar radius spread discussed above affect only one of the three estimators of the stellar radius. Hence, we would expect no correlation between the various estimators if there was no intrinsic spread in stellar radii. On the other hand an intrinsic spread of stellar radii should result in a clear correlation with brighter stars having a lower surface gravity and larger \(R \sin i\). The strongest correlation is found between the surface gravity and the extinction-corrected J with a Kendall's \(\tau\) \footnote{The Kendall's \(\tau\) ranked correlation is used here instead of the more commonly used Pearson r correlation, because of the sensitivity of Pearson r to outliers.} of 0.4, so that we can reject the null-hypothesis of no correlation (and hence no intrinsic spread in stellar radii) with a p-value much smaller than \(10^{-6}\). Weaker correlations are found between the \(R \sin i\) and the surface gravity (p-value of 3\%) or the extinction-corrected J (p-value of 0.5\%). The much lower significance in these correlations is caused both by the large scatter in the \(R \sin i\) values due to the multiplication with \(\sin i\) and the much smaller sample size. Visually the scatter plots in Figure \ref{fig:triangle} illustrate that these correlations nicely line up with the correlation expected for an intrinsic spread in stellar radii (as illustrated by the red contours).

Interpreting this correlation as a spread in stellar radii requires assuming that the other contributors to the spread in these stellar parameters are uncorrelated. Although this is true for the contributors discussed above, all estimators depend on the effective temperature, which could cause a spurious correlation. We took out any trend with the actual effective temperature of the star, by subtracting the empirical temperature-dependent trend line from all three stellar size estimators. However, this subtraction could cause a correlation, because an offset in the observed effective temperature would alter in a systematic way the trend-line value we subtract for all three stellar-size estimators. From the slope of the trend line we can estimate the size of this effect. The arrows in Figure \ref{fig:triangle} show the how a star would shift in these diagrams, if its actual effective temperature was about 100 K higher than measured by the pipeline. This offset of 100 K was chosen because it represents the maximum systematic offsets expected due to the clustering of observed effective temperatures around every other model grid point discussed in Section \ref{sec:Teff} and is significantly larger than the typical uncertainties in the effective temperature (Figure \ref{fig:dist_unc}). These arrows illustrate that in all but one (i.e. extinction-corrected J vs. \(R \sin i\)) case the expected correlation due to an offset in effective temperature is inconsistent with the direction of the observed correlation. Furthermore the size of the effect from these effective temperature offsets is very small compared with the total scatter, so we conclude that systematic or random effective temperature offsets have little to no effect on the stellar radius distribution measured here.

\subsection{Potential age spread}
\label{sec-5-2}
\label{sec:age_spread}
The most straight-forward explanation of the intrinsic stellar radius spread in IC 348 is that it corresponds to a spread in stellar ages, with younger stars still having larger radii, because they have had less time to contract towards the main sequence. The exact age spread that the stellar radius spread of 25\% (FWHM of 70\%) corresponds to depends both on the pre-main sequence evolutionary models that are adopted as well as the median absolute age of IC 348. Of our three estimates of the stellar radius, the trend of the extinction-corrected magnitude seems to be most consistent with the \citet{Dotter08} isochrones, with a median age of about 6 Myr. This age matches the recent estimate from \citet{Bell13a}, although it is significantly older than previous estimates \citep[e.g.][]{Luhman03}. For this median age the FWHM in the stellar radius spread would correspond to an age spread between 2 and 10 Myr. If IC 348 were younger, the age spread implied by the stellar radius spread would also be smaller (e.g., 1-5 Myr for a median age of 3 Myr). This age spread is consistent with models of star-formation, which lasts for several free-fall times of the molecular clouds \citep{Tan06, Nakamura07}. However, the stellar radius spread can be caused by other effects other than just a difference in contraction ages, so we can not exclude that the actual age spread is much smaller, as suggested by models of rapid star-formation \citep{Elmegreen00, Elmegreen07, Hartmann07}. Here we will look at such alternative scenarios that might explain the spread in stellar radii.

\begin{figure}[htb]
\centering
\includegraphics[width=.9\linewidth]{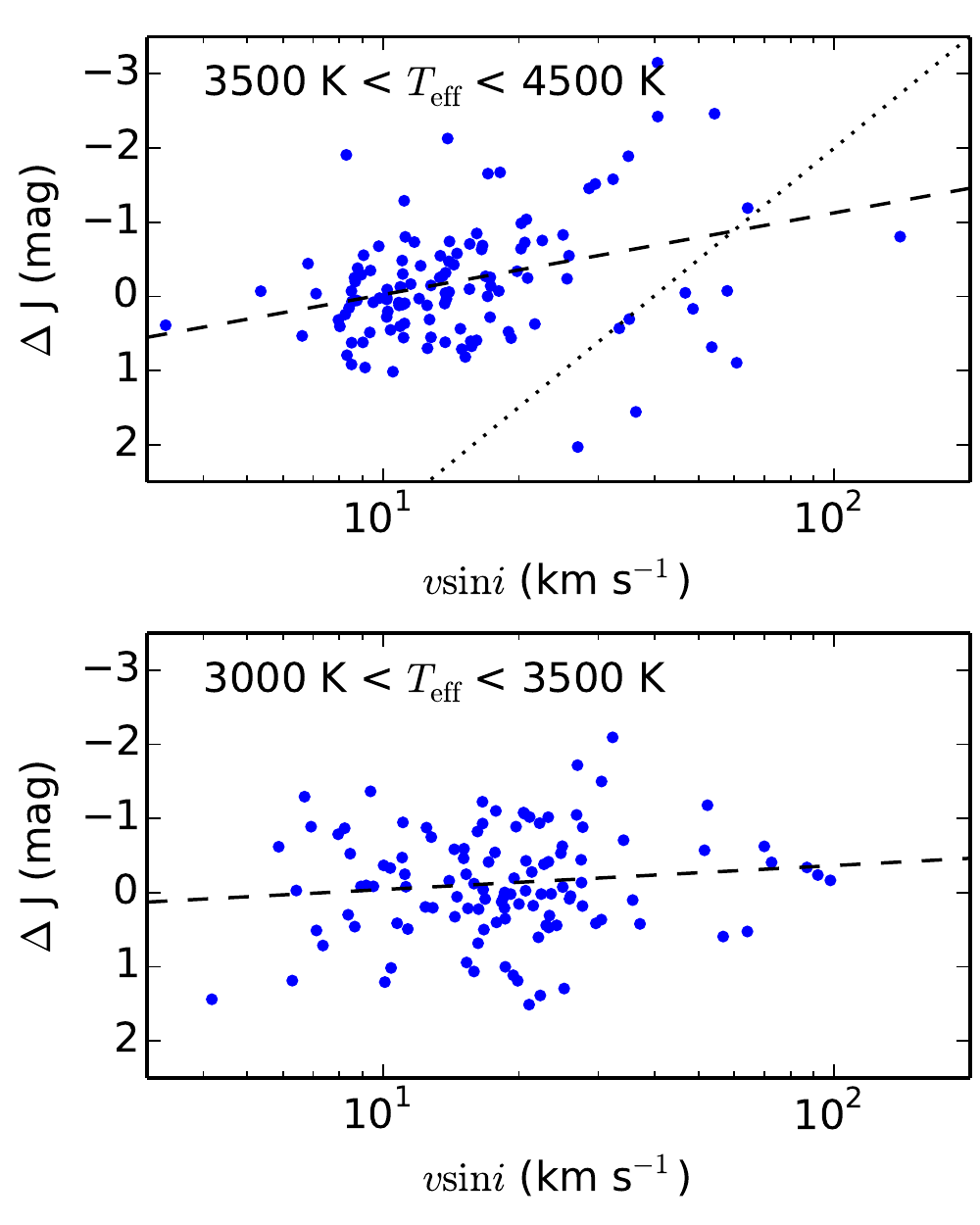}
\caption{\label{fig:rot_corr}Illustration of the correlation between the rotational velocity and the extinction-corrected J-band magnitude, where the trend with effective temperature has been subtracted from the latter. The upper panel contains stars with effective temperatures between 3500 and 4500 K and the lower panel stars between 3000 and 3500 K. A linear fit has been overplotted to guide the eye. The dotted line shows a constant period.}
\end{figure}

These alternative scenarios play at least some role, because the age-spread scenario described above can not explain our finding that the scatter of the stars in \(\Delta J\) is positively correlated with the stellar rotation rates with high statistical significance. In this analysis we include all stars with temperatures in the range 3000--4500 K, for which our empirical corrections are valid (see Figure \ref{fig:Teff_logg}). The \(v \sin i\) measures are used instead of the literature rotation periods because the latter are only available for a small subset of the stars\footnote{In this analysis we include the slow rotators, whose rotational velocity is sensitive to the resolution of the spectra, although we excluded them when analyzing the \(R \sin i\) distribution. We made this choice, because the total range of \(v \sin i\)-values is much larger than that for the \(R \sin i\)-values, which reduces the significance of the systematic offsets in the \(v \sin i\) due to the spectral resolution.}. We furthermore exclude nine stars with \(\Delta J > 3\) as outliers. The null hypothesis of no correlation between \(\Delta J\) and \(v \sin i\) is excluded with 99.8\% confidence using the nonparametric Kendall's \(\tau\) rank correlation test, strongly suggesting a correlation between the two quantities. If we only include K-stars (\(T_{\rm eff} > 3500\) K) the trend is even stronger, with a significance of 99.99\% in the Kendall's \(\tau\) rank correlation test (see Figure \ref{fig:rot_corr}). The sense of the correlation is of \(\Delta J\) becoming more negative (brighter) with increasing \(v \sin i\), so more rapidly rotating stars are larger. The larger radius for rapid rotators is expected to reduce the rate of Lithium burning, which could explain the larger Lithium abundances observed for more rapid rotators in the Pleiades \citep{Somers14}.

This correlation cannot be explained if the intrinsic stellar radius spread is fully caused by an age spread. A star that is decoupled from its circumstellar disc would spin up during contraction due to conservation of angular momentum, which causes a trend opposite to the one observed. A star that is tidally locked to the inner disc could lose angular momentum to the disc during contraction. However, the trend line in Figure \ref{fig:rot_corr} corresponds to \(v \sin i \propto R^{16}\), implying that a huge amount of angular momentum should be lost during contraction to explain this trend. Instead, to first order (i.e., assuming a constant magnetic field strength and accretion rate), we expect even tidally-locked stars to spin up during contraction due to a decrease in the inner disc radius to which the star is tidally locked (see equation 10.70 in \citealt{Hartmann09}). This analysis does not imply that the stars do not spin up during contraction; instead we argue that contraction across the small range of observed radii (total spread of 70\%) is unlikely to change its \(v \sin i\) sufficiently to broaden the large spread observed in \(v \sin i\) and hence the \(v \sin i\) distribution can not be used to constrain the physics of this contraction.

Such a trend between extinction-corrected stellar brightness and rotational velocity has also been observed by \citet{Littlefair11} and \citet{Kamaipress}. Based on simulations from \citet{Baraffe09}, \citet{Littlefair11} argue that this trend is likely caused by a difference in accretion history, for which they suggest several scenarios in which the accretion history sets both the current stellar radius and the stellar radius at the time the star decoupled from the disc. The smaller the star is at the time of decoupling, the smaller the subsequent spin-up will be, which leads to a more slowly rotating star. However, later studies have raised doubts about whether changes in accretion history can create a significant apparent age spread \citep{Hosokawa11, Baraffe12}. \citet{Hosokawa11} find that a variation in initial conditions (i.e. initial radius and thermal efficiency with which the accretion energy is absorbed) can cause an initial age spread. Interestingly they only found that variations in initial conditions formed an apparent age spread for hot stars (\(T_{\rm eff} > 3500\) K), which matches the range over which we find a correlation between stellar radius and rotation rate (Figure \ref{fig:rot_corr}). Alternatively a number of studies have found that magnetic activity among low-mass stars causes a suppression of the stellar surface temperature by of order 5--10\% and accordingly an inflation of the stellar radius by of order 10--20\%. This effect has been demonstrated in studies of low-mass eclipsing binaries as well as low-mass K and M dwarfs in the field \citep[e.g.,][]{Lopez-Morales07,Morales08,Stassun12}. Therefore, we may expect that the young, low-mass stars in our IC 348 sample, which possess a range of observed rotation rates, should also possess a range of magnetic activity levels and therefore a range of radius inflation. However, \citet{Alexander12} found no correlation between magnetic activity (as measured through the X-ray flux) and stellar rotation in IC 348, in line with similar measurements of T-Tauri stars in Orion and Taurus \citep{Preibisch05, Briggs07}.
\section{Summary}
\label{sec-6}
\label{sec:conc}
In this paper we present the spectral and photometric analysis of the young stars observed by SDSS-III/APOGEE as part of the "INfrared Spectra of Young Nebulous Clusters" (IN-SYNC) program. The spectra were fitted with BT-Settl model spectra using a forward modeling approach. Using epoch-to-epoch variability we showed that the true variability was about \(3 \sqrt{\chi^2_{\rm red,\ cont}}\) larger than estimated from the error array of the spectra using a Markov chain Monte Carlo (MCMC) simulation (equation \ref{eq:sigtrue}), where the reduced \(\chi^2\) was corrected for the systematic offset between model and observed spectrum (equation \ref{eq:chicorr}), so that it reflects how well the uncertainty in the observed flux has been estimated. The resulting estimates of the precision in the stellar parameters have a broad distribution (see Figure \ref{fig:dist_unc}). We note that for most stars the uncertainties in the radial velocity are significantly below the expected velocity dispersions in these young star-forming regions (\(\sim 1\) km s\(^{-1}\)), which implies that the observed velocity distribution should match the intrinsic radial velocity distribution in these regions (e.g., Cottaar et al., in prep). The resulting uncertainty distribution in the spectral parameters has strong non-Gaussian wings and is well described by a Cauchy-Lorentzian distribution with a width of 0.6 times the quoted measurement uncertainty (equation \ref{eq:Cauchy}).

In addition, we characterized in detail various systematic uncertainties in the spectral parameters:
\begin{enumerate}
\item We find and correct for a systematic offset in the radial velocity for the coolest stars (\(T_{\rm eff} < 3500\) K). We speculate that this offset might be caused by the line list, because the same offset is found for multiple model grids (both the BT-Settl grid from \citealt{Allard11} and the Gaia-ESO grid from \citealt{Husser13}), for H-band spectra from multiple instruments (both the APOGEE spectra and the NIRSPEC spectra from \citealt{Prato07}), and for multiple algorithms to measure the radial velocity (both the forward-modeling approach used here and the cross-correlation used in the standard APOGEE radial velocity pipeline).
\item A systematic offset of up to 400 K is found between the effective temperatures quoted here and the effective temperature scale from \citet{Luhman99}. After correcting for this offset only a small scatter (\(\sim 80\) K) remains between the effective temperatures measured here and those in the literature for cool stars (\(T_{\rm eff} < 3800\) K), which is likely caused a noding effect in the effective temperature, where the best fits tend to cluster around every other effective temperature in the model grid (Figure \ref{fig:Teff_logg}). For hotter stars this scatter is larger. Especially in the younger clusters with ages below 10 Myr these offsets can go up to several hundreds of Kelvin.
\item For cool stars (\(T_{\rm eff} < 4500\) K) the surface gravity is precise enough to distinguish the various young clusters observed here, as well as resolve a spread in stellar radii observed in IC 348. There are temperature-dependent systematic offsets with the Dartmouth isochrones, which become especially large for the hotter (\(T_{\rm eff} > 4500\) K) stars in the Pleiades. At these temperatures the surface gravities of the stars in the younger clusters is very large, which suggests the surface gravities of these hot stars should be treated with caution.
\item The rotational velocities are systematically higher than the literature values and the \(R \sin i\) calculated for those stars with derived stellar periods are also systematically higher than the stellar radii derived from either the surface gravities or the extinction-corrected luminosities. This overestimate of the rotational velocities is probably partly because of inaccuracies in the adopted instrumental line profile for APOGEE and partly because of the adopted profile for rotational broadening not being optimized for the near-infrared, where the limb broadening is lower than in the visible.
\end{enumerate}

Exploiting the uniform spectral and photometric dataset between the Pleiades and the young clusters observed as part of IN-SYNC we derive the stellar extinction from the J-H color excess of stars in the young cluster with respect to a single-star photometric locus of \citet{Bell12a}. Despite being based only on the near-infrared J-H color, the extinction estimates are accurate enough to closely match the excesses in the visible photometry from \citet{Bell12a, Bell13a} as well. Comparisons with these visible excesses suggest an uncertainty of 0.16 mag in \(A_{\rm J}\). 

Finally we combine the extinction-corrected luminosities, the surface gravities, and the \(R \sin i\) from the stellar rotation to show that IC 348 has a spread in stellar radii for stars between 3000 and 4500 K of about 25\%. By combining all three (nearly) independent estimates of the stellar radius we find that this spread is real and not caused by confusing hidden parameters, such as noise and binarity by showing that brighter stars tend to have significantly lower surface gravities and larger \(R \sin i\). This spread in stellar radii might correspond to an age spread of up to 8 Myr, depending on the absolute age of IC 348. However, we find that more rapid rotators tend to have larger stellar radii, which suggests that not all of the stellar radius spread is caused by a difference in contraction ages, but that at least some of the spread is caused by differences in accretion history or magnetic activity.
\section{Acknowledgements}
\label{sec-7}
MC ran simulations to develop the observational strategy for this program, developed and utilized the spectral analysis routines applied in this paper, led the scientific analysis of the stellar parameters, synthesized these results, and wrote the manuscript.  KRC, JCT and MRM conceived the program's scientific motivation and scope, led the initial ancillary science proposal, oversaw the project's progress and contributed to the analysis of the stellar parameters; KRC also led the target selection and sample design process, and provided assistance with the analysis and interpretation of the APOGEE spectra.  DLN assisted in the interpretation of the APOGEE data products and reduction algorithms, particularly those related to radial velocity measurements. JBF assisted with the analysis of the sample's completeness and extinction estimates. NDR selected the targets in Orion A. KMF assisted with target selection.  KGS contributed to the analysis of the luminosity spread presented in sections 5.1 and 5.2; SDC and GZ oversaw the design of the APOGEE plates utilized for IN-SYNC observations. SM, MS, and MFS contributed to defining the scope and implementation plan for this project, and with JCW and PMF developed and provided high level leadership for the broader APOGEE infrastructure that enabled this science. Furthermore we thank France Allard, Stella Offner, Rob Jeffries, Jinyoung Serena Kim, and Richard J. Parker for helpful comments and suggestions and thank Gus Muench and Luisa Rebull for help in the target selection of respectively IC 348 and NGC 1333.

MC and MRM acknowledge support from the Swiss National Science Foundation (SNF). Funding for SDSS-III has been provided by the Alfred P. Sloan Foundation, the Participating Institutions, the National Science Foundation, and the U.S. Department of Energy Office of Science. The SDSS-III web site is \url{http://www.sdss3.org/}.

SDSS-III is managed by the Astrophysical Research Consortium for the Participating Institutions of the SDSS-III Collaboration including the University of Arizona, the Brazilian Participation Group, Brookhaven National Laboratory, Carnegie Mellon University, University of Florida, the French Participation Group, the German Participation Group, Harvard University, the Instituto de Astrofisica de Canarias, the Michigan State/Notre Dame/JINA Participation Group, Johns Hopkins University, Lawrence Berkeley National Laboratory, Max Planck Institute for Astrophysics, Max Planck Institute for Extraterrestrial Physics, New Mexico State University, New York University, Ohio State University, Pennsylvania State University, University of Portsmouth, Princeton University, the Spanish Participation Group, University of Tokyo, University of Utah, Vanderbilt University, University of Virginia, University of Washington, and Yale University. PMF acknowledges support for this research from the National Science Foundation (AST-1311835).
\bibliographystyle{aa}
\bibliography{global}
\appendix
\section{Description of online tables}
\label{sec-8}
\label{app:tab}
Online we provide two companion tables to this paper, which contain the derived stellar parameters for the stars in IC 348 and the Pleiades. The first table contains one row per star with the mean spectral and photometric parameters. The second table contains one row per epoch with the spectral parameters measured at that epoch. In both tables we provide the uncertainties computed by equation \ref{eq:sigtrue}. Here we explain the meanings of all columns with in braces the names of the columns in the tables.

The first table contains 24 columns, starting with the 2MASS identifier (2MASS) and the right ascension (RA(deg)) and declination (Dec(deg)) in degrees. We also include which cluster the star belongs to or whether it is a field star observed on the same plate as these clusters (Cluster). This table then summarizes the observations with the number of epochs that spectra were taken (N(epochs)) with the total baseline in days (baseline(days)), and the S/N of a co-added spectrum (as derived from the spectrum error array; S/N). For all stellar parameters we report the weighted mean and its uncertainty over all epochs, as well as the probability that the stellar parameter is consistent with being constant, as estimated from the p-value that the \(\chi^2 = \frac{(p_i - \mu)^2}{\sigma_i^2}\) is larger than expected from chance. In this equation \(p_i\) is the measured parameter with uncertainty \(\sigma_i\) in epoch \(i\) and \(\mu\) is the weighted mean over all epochs. So we get the weighted mean of the effective temperature in Kelvin (Teff), its uncertainty (sig\_Teff), the p-value of a constant effective temperature (P(cnst\_Teff)), the weighted mean of the logarithmic surface gravity (log(g)), its uncertainty in dex (sig\_log(g)), the p-value of a constant surface gravity (P(cnst\_log(g))), weighted mean of the projected rotation velocity (vsini) in km s\(^{-1}\), its uncertainty (sig\_vsini), the p-value of a constant \(v \sin i\) (P(cnst\_vsini)), and finally the weighted mean of the veiling (R\_H), its uncertainty (sig\_R\_H), and the p-value for a constant veiling (P(cnst\_R\_H)). We also include the J (2MASS\_J), H (2MASS\_H), and Ks (2MASS\_Ks) 2MASS photometry, together with the J-H color excess (E(J-H)) and finally the extinction-corrected apparent magnitude \(m_{\rm J}\) (extinction-corrected\_J).

The second table contains for every epoch 15 columns, starting again with the 2MASS identifier (2MASS), which can be used to cross-link this table to the first table. In addition we include date of the observations in Julian days (date), the signal to noise ratio according to the error array of the observed spectrum (S/N), the effective temperature (Teff), its uncertainty in Kelvin (sig\_Teff), the normalized offset of the effective temperature from the weighted mean from the other spectra observed for this star (\(\eta_i\) from equation \ref{eq:eta}; eta\_Teff), the surface gravity (log(g)), its uncertainty in dex (sig\_log(g)), the \(\eta\)-offset from the mean surface gravity (eta\_log(g)), the rotational velocity in km s\(^{-1}\) (vsini), its uncertainty (sig\_vsini), the \(\eta\)-offset from the mean rotational velocity the uncorrected radial velocity (eta\_vsini), and finally the H-band veiling (R\_H), its uncertainty (sig\_R\_H), and the \(\eta\)-offset from the mean stellar veiling (eta\_R\_H).
\end{document}